\documentclass[prl,a4paper,secnumarabic,twocolumn,nofootinbib,nobibnotes,showpacs,amsmath,amssymb,tbtags,italian,frenchb,german,latin,swedish,british%,draft,%eqsecnum
]{revtex4}
\usepackage[T1]{fontenc} 
\usepackage{textcomp}
\renewenvironment{acknowledgements}{\section*{Acknowledgements}}{\par}
\usepackage{babel} 
\newcommand{\langfrench}{\foreignlanguage{french}}
\newcommand{\langgerman}{\foreignlanguage{german}}
\newcommand{\langitalian}{\foreignlanguage{italian}}

\newcommand{\french}[1]{{\langfrench{#1}}}
\usepackage{amsthm}
\usepackage[dvipdf]{graphicx}

\usepackage{url}
\usepackage{txfonts}
\newcommand{\etxtgr}[1][1]{
\DeclareFontFamily{U}{egreek}{\skewchar\font'177}%
\DeclareFontShape{U}{egreek}{m}{n}{<-6>s*[#1]eurm5<6-8>s*[#1]eurm7 <8->s*[#1]eurm10}{}%
\DeclareFontShape{U}{egreek}{m}{it}{<->s*[#1]eurmo10}{}%
\DeclareFontShape{U}{egreek}{b}{n}{<-6>s*[#1]eurb5 <6-8>s*[#1]eurb7 <8->s*[#1]eurb10}{}%
\DeclareFontShape{U}{egreek}{b}{it}{<->s*[#1]eurbo10}{}
}
\etxtgr
\newcommand{\partialup}{\text{\usefont{U}{egreek}{m}{n}@}}

\newcommand{\pgreeki}[1][.89]{%
\DeclareFontFamily{U}{pgreek}{}
\DeclareFontShape{U}{pgreek}{m}{n}{<->s*[#1]psyr}{}
\DeclareFontShape{U}{pgreek}{m}{it}{<->s*[#1]psyro}{}
\DeclareSymbolFont{pgreeki}{U}{pgreek}{m}{it}
\DeclareSymbolFontAlphabet{\mathpgri}{pgreeki}%
\DeclareMathSymbol{\palpha}{\mathalpha}{pgreeki}{`a}
\DeclareMathSymbol{\pbeta}{\mathalpha}{pgreeki}{`b}
\DeclareMathSymbol{\pgamma}{\mathalpha}{pgreeki}{`g}
\DeclareMathSymbol{\pdelta}{\mathalpha}{pgreeki}{`d}
\DeclareMathSymbol{\pepsilon}{\mathalpha}{pgreeki}{`e}
\DeclareMathSymbol{\pzeta}{\mathalpha}{pgreeki}{`z}
\DeclareMathSymbol{\peta}{\mathalpha}{pgreeki}{`h}
\DeclareMathSymbol{\ptheta}{\mathalpha}{pgreeki}{`q}
\DeclareMathSymbol{\piota}{\mathalpha}{pgreeki}{`i}
\DeclareMathSymbol{\pkappa}{\mathalpha}{pgreeki}{`k}
\DeclareMathSymbol{\plambda}{\mathalpha}{pgreeki}{`l}
\DeclareMathSymbol{\pmu}{\mathalpha}{pgreeki}{`m}
\DeclareMathSymbol{\pnu}{\mathalpha}{pgreeki}{`n}
\DeclareMathSymbol{\pxi}{\mathalpha}{pgreeki}{`x}
\DeclareMathSymbol{\pomicron}{\mathalpha}{pgreeki}{`o}
\DeclareMathSymbol{\ppi}{\mathalpha}{pgreeki}{`p}
\DeclareMathSymbol{\prho}{\mathalpha}{pgreeki}{`r}
\DeclareMathSymbol{\psigma}{\mathalpha}{pgreeki}{`s}
\DeclareMathSymbol{\ptau}{\mathalpha}{pgreeki}{`t}
\DeclareMathSymbol{\pupsilon}{\mathalpha}{pgreeki}{`u}
\DeclareMathSymbol{\pphi}{\mathalpha}{pgreeki}{`f}
\DeclareMathSymbol{\pchi}{\mathalpha}{pgreeki}{`c}
\DeclareMathSymbol{\ppsi}{\mathalpha}{pgreeki}{`y}
\DeclareMathSymbol{\pomega}{\mathalpha}{pgreeki}{`w}
\let\pvarepsilon\pepsilon
\DeclareMathSymbol{\pvartheta}{\mathalpha}{pgreeki}{`J}
\DeclareMathSymbol{\pvarpi}{\mathalpha}{pgreeki}{`v}
\let\pvarrho\prho 
\let\pvarkappa\pkappa 
\DeclareMathSymbol{\pvarsigma}{\mathalpha}{pgreeki}{`V}
\DeclareMathSymbol{\pvarphi}{\mathalpha}{pgreeki}{`j}
\DeclareMathSymbol{\pvarAlpha}{\mathalpha}{pgreeki}{`A}
\DeclareMathSymbol{\pvarBeta}{\mathalpha}{pgreeki}{`B}
\DeclareMathSymbol{\pvarGamma}{\mathalpha}{pgreeki}{`G}
\DeclareMathSymbol{\pvarDelta}{\mathalpha}{pgreeki}{`D}
\DeclareMathSymbol{\pvarEpsilon}{\mathalpha}{pgreeki}{`E}
\DeclareMathSymbol{\pvarZeta}{\mathalpha}{pgreeki}{`Z}
\DeclareMathSymbol{\pvarEta}{\mathalpha}{pgreeki}{`H}
\DeclareMathSymbol{\pvarTheta}{\mathalpha}{pgreeki}{`Q}
\DeclareMathSymbol{\pvarIota}{\mathalpha}{pgreeki}{`I}
\DeclareMathSymbol{\pvarKappa}{\mathalpha}{pgreeki}{`K}
\DeclareMathSymbol{\pvarLambda}{\mathalpha}{pgreeki}{`L}
\DeclareMathSymbol{\pvarMu}{\mathalpha}{pgreeki}{`M}
\DeclareMathSymbol{\pvarNu}{\mathalpha}{pgreeki}{`N}
\DeclareMathSymbol{\pvarXi}{\mathalpha}{pgreeki}{`X}
\DeclareMathSymbol{\pvarOmicron}{\mathalpha}{pgreeki}{`O}
\DeclareMathSymbol{\pvarPi}{\mathalpha}{pgreeki}{`P}
\DeclareMathSymbol{\pvarRho}{\mathalpha}{pgreeki}{`R}
\DeclareMathSymbol{\pvarSigma}{\mathalpha}{pgreeki}{`S}
\DeclareMathSymbol{\pvarTau}{\mathalpha}{pgreeki}{`T}
\DeclareMathSymbol{\pvarUpsilon}{\mathalpha}{pgreeki}{`U}
\DeclareMathSymbol{\pvarPhi}{\mathalpha}{pgreeki}{`F}
\DeclareMathSymbol{\pvarChi}{\mathalpha}{pgreeki}{`C}
\DeclareMathSymbol{\pvarPsi}{\mathalpha}{pgreeki}{`Y}
\DeclareMathSymbol{\pvarOmega}{\mathalpha}{pgreeki}{`W} 
}

\newcommand{\pgreekr}[1][.89]{%
\DeclareFontFamily{U}{pgreek}{}
\DeclareFontShape{U}{pgreek}{m}{n}{<->s*[#1]psyr}{}
\DeclareFontShape{U}{pgreek}{m}{it}{<->s*[#1]psyro}{}
\DeclareSymbolFont{pgreekr}{U}{pgreek}{m}{n}
\DeclareSymbolFontAlphabet{\mathpgrr}{pgreekr}%
\DeclareMathSymbol{\palphaup}{\mathalpha}{pgreekr}{`a}
\DeclareMathSymbol{\pbetaup}{\mathalpha}{pgreekr}{`b}
\DeclareMathSymbol{\pgammaup}{\mathalpha}{pgreekr}{`g}
\DeclareMathSymbol{\pdeltaup}{\mathalpha}{pgreekr}{`d}
\DeclareMathSymbol{\pepsilonup}{\mathalpha}{pgreekr}{`e}
\DeclareMathSymbol{\pzetaup}{\mathalpha}{pgreekr}{`z}
\DeclareMathSymbol{\petaup}{\mathalpha}{pgreekr}{`h}
\DeclareMathSymbol{\pthetaup}{\mathalpha}{pgreekr}{`q}
\DeclareMathSymbol{\piotaup}{\mathalpha}{pgreekr}{`i}
\DeclareMathSymbol{\pkappaup}{\mathalpha}{pgreekr}{`k}
\DeclareMathSymbol{\plambdaup}{\mathalpha}{pgreekr}{`l}
\DeclareMathSymbol{\pmuup}{\mathalpha}{pgreekr}{`m}
\DeclareMathSymbol{\pnuup}{\mathalpha}{pgreekr}{`n}
\DeclareMathSymbol{\pxiup}{\mathalpha}{pgreekr}{`x}
\DeclareMathSymbol{\pomicronup}{\mathalpha}{pgreekr}{`o}
\DeclareMathSymbol{\ppiup}{\mathalpha}{pgreekr}{`p}
\DeclareMathSymbol{\prhoup}{\mathalpha}{pgreekr}{`r}
\DeclareMathSymbol{\psigmaup}{\mathalpha}{pgreekr}{`s}
\DeclareMathSymbol{\ptauup}{\mathalpha}{pgreekr}{`t}
\DeclareMathSymbol{\pupsilonup}{\mathalpha}{pgreekr}{`u}
\DeclareMathSymbol{\pphiup}{\mathalpha}{pgreekr}{`f}
\DeclareMathSymbol{\pchiup}{\mathalpha}{pgreekr}{`c}
\DeclareMathSymbol{\ppsiup}{\mathalpha}{pgreekr}{`y}
\DeclareMathSymbol{\pomegaup}{\mathalpha}{pgreekr}{`w}
\let\pvarepsilonup\pepsilonup
\DeclareMathSymbol{\pvarthetaup}{\mathalpha}{pgreekr}{`J}
\DeclareMathSymbol{\pvarpiup}{\mathalpha}{pgreekr}{`v}
\let\pvarrhoup\prhoup 
\DeclareMathSymbol{\pvarsigmaup}{\mathalpha}{pgreekr}{`V}
\DeclareMathSymbol{\pvarphiup}{\mathalpha}{pgreekr}{`j}
\let\pvarkappaup\pkappaup 
\DeclareMathSymbol{\pAlpha}{\mathalpha}{pgreekr}{`A}
\DeclareMathSymbol{\pBeta}{\mathalpha}{pgreekr}{`B}
\DeclareMathSymbol{\pGamma}{\mathalpha}{pgreekr}{`G}
\DeclareMathSymbol{\pDelta}{\mathalpha}{pgreekr}{`D}
\DeclareMathSymbol{\pEpsilon}{\mathalpha}{pgreekr}{`E}
\DeclareMathSymbol{\pZeta}{\mathalpha}{pgreekr}{`Z}
\DeclareMathSymbol{\pEta}{\mathalpha}{pgreekr}{`H}
\DeclareMathSymbol{\pTheta}{\mathalpha}{pgreekr}{`Q}
\DeclareMathSymbol{\pIota}{\mathalpha}{pgreekr}{`I}
\DeclareMathSymbol{\pKappa}{\mathalpha}{pgreekr}{`K}
\DeclareMathSymbol{\pLambda}{\mathalpha}{pgreekr}{`L}
\DeclareMathSymbol{\pMu}{\mathalpha}{pgreekr}{`M}
\DeclareMathSymbol{\pNu}{\mathalpha}{pgreekr}{`N}
\DeclareMathSymbol{\pXi}{\mathalpha}{pgreekr}{`X}
\DeclareMathSymbol{\pOmicron}{\mathalpha}{pgreekr}{`O}
\DeclareMathSymbol{\pPi}{\mathalpha}{pgreekr}{`P}
\DeclareMathSymbol{\pRho}{\mathalpha}{pgreekr}{`R}
\DeclareMathSymbol{\pSigma}{\mathalpha}{pgreekr}{`S}
\DeclareMathSymbol{\pTau}{\mathalpha}{pgreekr}{`T}
\DeclareMathSymbol{\pUpsilon}{\mathalpha}{pgreekr}{`U}
\DeclareMathSymbol{\pPhi}{\mathalpha}{pgreekr}{`F}
\DeclareMathSymbol{\pChi}{\mathalpha}{pgreekr}{`C}
\DeclareMathSymbol{\pPsi}{\mathalpha}{pgreekr}{`Y}
\DeclareMathSymbol{\pOmega}{\mathalpha}{pgreekr}{`W} 
}

\newcommand{\egreeki}[1][1]{%
\DeclareFontFamily{U}{egreek}{\skewchar\font'177}%
\DeclareFontShape{U}{egreek}{m}{n}{<-6>s*[#1]eurm5<6-8>s*[#1]eurm7 <8->s*[#1]eurm10}{}%
\DeclareFontShape{U}{egreek}{m}{it}{<->s*[#1]eurmo10}{}%
\DeclareFontShape{U}{egreek}{b}{n}{<-6>s*[#1]eurb5 <6-8>s*[#1]eurb7 <8->s*[#1]eurb10}{}%
\DeclareFontShape{U}{egreek}{b}{it}{<->s*[#1]eurbo10}{}%
\DeclareSymbolFont{egreeki}{U}{egreek}{m}{it}%
\SetSymbolFont{egreeki}{bold}{U}{egreek}{b}{it}% from the amsfonts package
\DeclareSymbolFontAlphabet{\mathegri}{egreeki}%
\DeclareMathSymbol{\epartial}{\mathalpha}{egreeki}{"40}
\DeclareMathSymbol{\ealpha}{\mathalpha}{egreeki}{"0B}
\DeclareMathSymbol{\ebeta}{\mathalpha}{egreeki}{"0C}
\DeclareMathSymbol{\egamma}{\mathalpha}{egreeki}{"0D}
\DeclareMathSymbol{\edelta}{\mathalpha}{egreeki}{"0E}
\DeclareMathSymbol{\eepsilon}{\mathalpha}{egreeki}{"0F}
\DeclareMathSymbol{\ezeta}{\mathalpha}{egreeki}{"10}
\DeclareMathSymbol{\eeta}{\mathalpha}{egreeki}{"11}
\DeclareMathSymbol{\etheta}{\mathalpha}{egreeki}{"12}
\DeclareMathSymbol{\eiota}{\mathalpha}{egreeki}{"13}
\DeclareMathSymbol{\ekappa}{\mathalpha}{egreeki}{"14}
\DeclareMathSymbol{\elambda}{\mathalpha}{egreeki}{"15}
\DeclareMathSymbol{\emu}{\mathalpha}{egreeki}{"16}
\DeclareMathSymbol{\enu}{\mathalpha}{egreeki}{"17}
\DeclareMathSymbol{\exi}{\mathalpha}{egreeki}{"18}
\DeclareMathSymbol{\eomicron}{\mathalpha}{egreeki}{"6F}
\DeclareMathSymbol{\epi}{\mathalpha}{egreeki}{"19}
\DeclareMathSymbol{\erho}{\mathalpha}{egreeki}{"1A}
\DeclareMathSymbol{\esigma}{\mathalpha}{egreeki}{"1B}
\DeclareMathSymbol{\etau}{\mathalpha}{egreeki}{"1C}
\DeclareMathSymbol{\eupsilon}{\mathalpha}{egreeki}{"1D}
\DeclareMathSymbol{\ephi}{\mathalpha}{egreeki}{"1E}
\DeclareMathSymbol{\echi}{\mathalpha}{egreeki}{"1F}
\DeclareMathSymbol{\epsi}{\mathalpha}{egreeki}{"20}
\DeclareMathSymbol{\eomega}{\mathalpha}{egreeki}{"21}
\DeclareMathSymbol{\evarepsilon}{\mathalpha}{egreeki}{"22}
\DeclareMathSymbol{\evartheta}{\mathalpha}{egreeki}{"23}
\DeclareMathSymbol{\evarpi}{\mathalpha}{egreeki}{"24}
\let\evarrho\erho 
\let\evarsigma\esigma
\let\evarkappa\ekappa
\DeclareMathSymbol{\evarphi}{\mathalpha}{egreeki}{"27}
\DeclareMathSymbol{\evarAlpha}{\mathalpha}{egreeki}{"41}
\DeclareMathSymbol{\evarBeta}{\mathalpha}{egreeki}{"42}
\DeclareMathSymbol{\evarGamma}{\mathalpha}{egreeki}{"00}
\DeclareMathSymbol{\evarDelta}{\mathalpha}{egreeki}{"01}
\DeclareMathSymbol{\evarEpsilon}{\mathalpha}{egreeki}{"45}
\DeclareMathSymbol{\evarZeta}{\mathalpha}{egreeki}{"5A}
\DeclareMathSymbol{\evarEta}{\mathalpha}{egreeki}{"48}
\DeclareMathSymbol{\evarTheta}{\mathalpha}{egreeki}{"02}
\DeclareMathSymbol{\evarIota}{\mathalpha}{egreeki}{"49}
\DeclareMathSymbol{\evarKappa}{\mathalpha}{egreeki}{"4B}
\DeclareMathSymbol{\evarLambda}{\mathalpha}{egreeki}{"03}
\DeclareMathSymbol{\evarMu}{\mathalpha}{egreeki}{"4D}
\DeclareMathSymbol{\evarNu}{\mathalpha}{egreeki}{"4E}
\DeclareMathSymbol{\evarXi}{\mathalpha}{egreeki}{"04}
\DeclareMathSymbol{\evarOmicron}{\mathalpha}{egreeki}{"4F}
\DeclareMathSymbol{\evarPi}{\mathalpha}{egreeki}{"05}
\DeclareMathSymbol{\evarRho}{\mathalpha}{egreeki}{"50}
\DeclareMathSymbol{\evarSigma}{\mathalpha}{egreeki}{"06}
\DeclareMathSymbol{\evarTau}{\mathalpha}{egreeki}{"54}
\DeclareMathSymbol{\evarUpsilon}{\mathalpha}{egreeki}{"07}
\DeclareMathSymbol{\evarPhi}{\mathalpha}{egreeki}{"08}
\DeclareMathSymbol{\evarChi}{\mathalpha}{egreeki}{"58}
\DeclareMathSymbol{\evarPsi}{\mathalpha}{egreeki}{"09}
\DeclareMathSymbol{\evarOmega}{\mathalpha}{egreeki}{"0A} 
}

\newcommand{\egreekr}[1][1]{%
\DeclareFontFamily{U}{egreek}{\skewchar\font'177}%
\DeclareFontShape{U}{egreek}{m}{n}{<-6>s*[#1]eurm5<6-8>s*[#1]eurm7 <8->s*[#1]eurm10}{}%
\DeclareFontShape{U}{egreek}{m}{it}{<->s*[#1]eurmo10}{}%
\DeclareFontShape{U}{egreek}{b}{n}{<-6>s*[#1]eurb5 <6-8>s*[#1]eurb7 <8->s*[#1]eurb10}{}%
\DeclareFontShape{U}{egreek}{b}{it}{<->s*[#1]eurbo10}{}%
\DeclareSymbolFont{egreekr}{U}{egreek}{m}{n}%
\SetSymbolFont{egreekr}{bold}{U}{egreek}{b}{n}% from the amsfonts package
\DeclareSymbolFontAlphabet{\mathegrr}{egreekr}%
\DeclareMathSymbol{\epartialup}{\mathalpha}{egreekr}{"40}
\let\partialup\epartialup
\DeclareMathSymbol{\ealphaup}{\mathalpha}{egreekr}{"0B}
\DeclareMathSymbol{\ebetaup}{\mathalpha}{egreekr}{"0C}
\DeclareMathSymbol{\egammaup}{\mathalpha}{egreekr}{"0D}
\DeclareMathSymbol{\edeltaup}{\mathalpha}{egreekr}{"0E}
\DeclareMathSymbol{\eepsilonup}{\mathalpha}{egreekr}{"0F}
\DeclareMathSymbol{\ezetaup}{\mathalpha}{egreekr}{"10}
\DeclareMathSymbol{\eetaup}{\mathalpha}{egreekr}{"11}
\DeclareMathSymbol{\ethetaup}{\mathalpha}{egreekr}{"12}
\DeclareMathSymbol{\eiotaup}{\mathalpha}{egreekr}{"13}
\DeclareMathSymbol{\ekappaup}{\mathalpha}{egreekr}{"14}
\DeclareMathSymbol{\elambdaup}{\mathalpha}{egreekr}{"15}
\DeclareMathSymbol{\emuup}{\mathalpha}{egreekr}{"16}
\DeclareMathSymbol{\enuup}{\mathalpha}{egreekr}{"17}
\DeclareMathSymbol{\exiup}{\mathalpha}{egreekr}{"18}
\DeclareMathSymbol{\eomicronup}{\mathalpha}{egreekr}{"6F}
\DeclareMathSymbol{\epiup}{\mathalpha}{egreekr}{"19}
\DeclareMathSymbol{\erhoup}{\mathalpha}{egreekr}{"1A}
\DeclareMathSymbol{\esigmaup}{\mathalpha}{egreekr}{"1B}
\DeclareMathSymbol{\etauup}{\mathalpha}{egreekr}{"1C}
\DeclareMathSymbol{\eupsilonup}{\mathalpha}{egreekr}{"1D}
\DeclareMathSymbol{\ephiup}{\mathalpha}{egreekr}{"1E}
\DeclareMathSymbol{\echiup}{\mathalpha}{egreekr}{"1F}
\DeclareMathSymbol{\epsiup}{\mathalpha}{egreekr}{"20}
\DeclareMathSymbol{\eomegaup}{\mathalpha}{egreekr}{"21}
\DeclareMathSymbol{\evarepsilonup}{\mathalpha}{egreekr}{"22}
\DeclareMathSymbol{\evarthetaup}{\mathalpha}{egreekr}{"23}
\DeclareMathSymbol{\evarpiup}{\mathalpha}{egreekr}{"24}
\let\evarrhoup\erhoup 
\let\evarsigmaup\esigmaup
\let\evarkappaup\ekappaup
\DeclareMathSymbol{\evarphiup}{\mathalpha}{egreekr}{"27}
\DeclareMathSymbol{\eAlpha}{\mathalpha}{egreekr}{"41}
\DeclareMathSymbol{\eBeta}{\mathalpha}{egreekr}{"42}
\DeclareMathSymbol{\eGamma}{\mathalpha}{egreekr}{"00}
\DeclareMathSymbol{\eDelta}{\mathalpha}{egreekr}{"01}
\DeclareMathSymbol{\eEpsilon}{\mathalpha}{egreekr}{"45}
\DeclareMathSymbol{\eZeta}{\mathalpha}{egreekr}{"5A}
\DeclareMathSymbol{\eEta}{\mathalpha}{egreekr}{"48}
\DeclareMathSymbol{\eTheta}{\mathalpha}{egreekr}{"02}
\DeclareMathSymbol{\eIota}{\mathalpha}{egreekr}{"49}
\DeclareMathSymbol{\eKappa}{\mathalpha}{egreekr}{"4B}
\DeclareMathSymbol{\eLambda}{\mathalpha}{egreekr}{"03}
\DeclareMathSymbol{\eMu}{\mathalpha}{egreekr}{"4D}
\DeclareMathSymbol{\eNu}{\mathalpha}{egreekr}{"4E}
\DeclareMathSymbol{\eXi}{\mathalpha}{egreekr}{"04}
\DeclareMathSymbol{\eOmicron}{\mathalpha}{egreekr}{"4F}
\DeclareMathSymbol{\ePi}{\mathalpha}{egreekr}{"05}
\DeclareMathSymbol{\eRho}{\mathalpha}{egreekr}{"50}
\DeclareMathSymbol{\eSigma}{\mathalpha}{egreekr}{"06}
\DeclareMathSymbol{\eTau}{\mathalpha}{egreekr}{"54}
\DeclareMathSymbol{\eUpsilon}{\mathalpha}{egreekr}{"07}
\DeclareMathSymbol{\ePhi}{\mathalpha}{egreekr}{"08}
\DeclareMathSymbol{\eChi}{\mathalpha}{egreekr}{"58}
\DeclareMathSymbol{\ePsi}{\mathalpha}{egreekr}{"09}
\DeclareMathSymbol{\eOmega}{\mathalpha}{egreekr}{"0A}
}

\newcommand{\tgreekr}[1][1]{%
\DeclareSymbolFont{tgreekr}{OML}{txmia}{m}{n}%
\SetSymbolFont{tgreekr}{bold}{OML}{txmia}{bx}{n}% from the amsfonts package
\DeclareSymbolFontAlphabet{\mathtgrr}{tgreekr}%
\DeclareMathSymbol{\talphaup}{\mathalpha}{tgreekr}{"0B}
\DeclareMathSymbol{\tbetaup}{\mathalpha}{tgreekr}{"0C}
\DeclareMathSymbol{\tgammaup}{\mathalpha}{tgreekr}{"0D}
\DeclareMathSymbol{\tdeltaup}{\mathalpha}{tgreekr}{"0E}
\DeclareMathSymbol{\tepsilonup}{\mathalpha}{tgreekr}{"0F}
\DeclareMathSymbol{\tzetaup}{\mathalpha}{tgreekr}{"10}
\DeclareMathSymbol{\tetaup}{\mathalpha}{tgreekr}{"11}
\DeclareMathSymbol{\tthetaup}{\mathalpha}{tgreekr}{"12}
\DeclareMathSymbol{\tiotaup}{\mathalpha}{tgreekr}{"13}
\DeclareMathSymbol{\tkappaup}{\mathalpha}{tgreekr}{"14}
\DeclareMathSymbol{\tlambdaup}{\mathalpha}{tgreekr}{"15}
\DeclareMathSymbol{\tmuup}{\mathalpha}{tgreekr}{"16}
\DeclareMathSymbol{\tnuup}{\mathalpha}{tgreekr}{"17}
\DeclareMathSymbol{\txiup}{\mathalpha}{tgreekr}{"18}
\DeclareMathSymbol{\tomicronup}{\mathalpha}{tgreekr}{"6F}
\DeclareMathSymbol{\tpiup}{\mathalpha}{tgreekr}{"19}
\DeclareMathSymbol{\trhoup}{\mathalpha}{tgreekr}{"1A}
\DeclareMathSymbol{\tsigmaup}{\mathalpha}{tgreekr}{"1B}
\DeclareMathSymbol{\ttauup}{\mathalpha}{tgreekr}{"1C}
\DeclareMathSymbol{\tupsilonup}{\mathalpha}{tgreekr}{"1D}
\DeclareMathSymbol{\tphiup}{\mathalpha}{tgreekr}{"1E}
\DeclareMathSymbol{\tchiup}{\mathalpha}{tgreekr}{"1F}
\DeclareMathSymbol{\tpsiup}{\mathalpha}{tgreekr}{"20}
\DeclareMathSymbol{\tomegaup}{\mathalpha}{tgreekr}{"21}
\DeclareMathSymbol{\tvarepsilonup}{\mathalpha}{tgreekr}{"22}
\DeclareMathSymbol{\tvarthetaup}{\mathalpha}{tgreekr}{"23}
\DeclareMathSymbol{\tvarpiup}{\mathalpha}{tgreekr}{"24}
\let\tvarrhoup\trhoup 
\let\tvarsigmaup\tsigmaup
\let\tvarkappaup\tkappaup
\DeclareMathSymbol{\tvarphiup}{\mathalpha}{tgreekr}{"27}
\DeclareMathSymbol{\tAlpha}{\mathalpha}{tgreekr}{"41}
\DeclareMathSymbol{\tBeta}{\mathalpha}{tgreekr}{"42}
\DeclareMathSymbol{\tGamma}{\mathalpha}{tgreekr}{"00}
\DeclareMathSymbol{\tDelta}{\mathalpha}{tgreekr}{"01}
\DeclareMathSymbol{\tEpsilon}{\mathalpha}{tgreekr}{"45}
\DeclareMathSymbol{\tZeta}{\mathalpha}{tgreekr}{"5A}
\DeclareMathSymbol{\tEta}{\mathalpha}{tgreekr}{"48}
\DeclareMathSymbol{\tTheta}{\mathalpha}{tgreekr}{"02}
\DeclareMathSymbol{\tIota}{\mathalpha}{tgreekr}{"49}
\DeclareMathSymbol{\tKappa}{\mathalpha}{tgreekr}{"4B}
\DeclareMathSymbol{\tLambda}{\mathalpha}{tgreekr}{"03}
\DeclareMathSymbol{\tMu}{\mathalpha}{tgreekr}{"4D}
\DeclareMathSymbol{\tNu}{\mathalpha}{tgreekr}{"4E}
\DeclareMathSymbol{\tXi}{\mathalpha}{tgreekr}{"04}
\DeclareMathSymbol{\tOmicron}{\mathalpha}{tgreekr}{"4F}
\DeclareMathSymbol{\tPi}{\mathalpha}{tgreekr}{"05}
\DeclareMathSymbol{\tRho}{\mathalpha}{tgreekr}{"50}
\DeclareMathSymbol{\tSigma}{\mathalpha}{tgreekr}{"06}
\DeclareMathSymbol{\tTau}{\mathalpha}{tgreekr}{"54}
\DeclareMathSymbol{\tUpsilon}{\mathalpha}{tgreekr}{"07}
\DeclareMathSymbol{\tPhi}{\mathalpha}{tgreekr}{"08}
\DeclareMathSymbol{\tChi}{\mathalpha}{tgreekr}{"58}
\DeclareMathSymbol{\tPsi}{\mathalpha}{tgreekr}{"09}
\DeclareMathSymbol{\tOmega}{\mathalpha}{tgreekr}{"0A}
}

\newcommand{\tgreeki}[1][1]{%
\DeclareSymbolFont{tgreeki}{OML}{txmi}{m}{it}%
\SetSymbolFont{tgreeki}{bold}{OML}{txmi}{bx}{it}% from the amsfonts package
\DeclareSymbolFontAlphabet{\mathtgri}{tgreeki}%
 \DeclareMathSymbol{\talpha}{\mathalpha}{tgreeki}{"0B}
 \DeclareMathSymbol{\tbeta}{\mathalpha}{tgreeki}{"0C}
 \DeclareMathSymbol{\tgamma}{\mathalpha}{tgreeki}{"0D}
 \DeclareMathSymbol{\tdelta}{\mathalpha}{tgreeki}{"0E}
 \DeclareMathSymbol{\tepsilon}{\mathalpha}{tgreeki}{"0F}
 \DeclareMathSymbol{\tzeta}{\mathalpha}{tgreeki}{"10}
 \DeclareMathSymbol{\teta}{\mathalpha}{tgreeki}{"11}
 \DeclareMathSymbol{\ttheta}{\mathalpha}{tgreeki}{"12}
 \DeclareMathSymbol{\tiota}{\mathalpha}{tgreeki}{"13}
 \DeclareMathSymbol{\tkappa}{\mathalpha}{tgreeki}{"14}
 \DeclareMathSymbol{\tlambda}{\mathalpha}{tgreeki}{"15}
 \DeclareMathSymbol{\tmu}{\mathalpha}{tgreeki}{"16}
 \DeclareMathSymbol{\tnu}{\mathalpha}{tgreeki}{"17}
 \DeclareMathSymbol{\txi}{\mathalpha}{tgreeki}{"18}
 \DeclareMathSymbol{\tomicron}{\mathalpha}{tgreeki}{"6F}
 \DeclareMathSymbol{\tpi}{\mathalpha}{tgreeki}{"19}
 \DeclareMathSymbol{\trho}{\mathalpha}{tgreeki}{"1A}
 \DeclareMathSymbol{\tsigma}{\mathalpha}{tgreeki}{"1B}
 \DeclareMathSymbol{\ttau}{\mathalpha}{tgreeki}{"1C}
 \DeclareMathSymbol{\tupsilon}{\mathalpha}{tgreeki}{"1D}
 \DeclareMathSymbol{\tphi}{\mathalpha}{tgreeki}{"1E}
 \DeclareMathSymbol{\tchi}{\mathalpha}{tgreeki}{"1F}
 \DeclareMathSymbol{\tpsi}{\mathalpha}{tgreeki}{"20}
 \DeclareMathSymbol{\tomega}{\mathalpha}{tgreeki}{"21}
 \DeclareMathSymbol{\tvarepsilon}{\mathalpha}{tgreeki}{"22}
 \DeclareMathSymbol{\tvartheta}{\mathalpha}{tgreeki}{"23}
 \DeclareMathSymbol{\tvarpi}{\mathalpha}{tgreeki}{"24}
 \let\tvarrho\trho 
 \let\tvarsigma\tsigma
 \let\tvarkappa\tkappa
 \DeclareMathSymbol{\tvarphi}{\mathalpha}{tgreeki}{"27}
 \DeclareMathSymbol{\tvarAlpha}{\mathalpha}{tgreeki}{"41}
 \DeclareMathSymbol{\tvarBeta}{\mathalpha}{tgreeki}{"42}
 \DeclareMathSymbol{\tvarGamma}{\mathalpha}{tgreeki}{"00}
 \DeclareMathSymbol{\tvarDelta}{\mathalpha}{tgreeki}{"01}
 \DeclareMathSymbol{\tvarEpsilon}{\mathalpha}{tgreeki}{"45}
 \DeclareMathSymbol{\tvarZeta}{\mathalpha}{tgreeki}{"5A}
 \DeclareMathSymbol{\tvarEta}{\mathalpha}{tgreeki}{"48}
 \DeclareMathSymbol{\tvarTheta}{\mathalpha}{tgreeki}{"02}
 \DeclareMathSymbol{\tvarIota}{\mathalpha}{tgreeki}{"49}
 \DeclareMathSymbol{\tvarKappa}{\mathalpha}{tgreeki}{"4B}
 \DeclareMathSymbol{\tvarLambda}{\mathalpha}{tgreeki}{"03}
 \DeclareMathSymbol{\tvarMu}{\mathalpha}{tgreeki}{"4D}
 \DeclareMathSymbol{\tvarNu}{\mathalpha}{tgreeki}{"4E}
 \DeclareMathSymbol{\tvarXi}{\mathalpha}{tgreeki}{"04}
 \DeclareMathSymbol{\tvarOmicron}{\mathalpha}{tgreeki}{"4F}
 \DeclareMathSymbol{\tvarPi}{\mathalpha}{tgreeki}{"05}
 \DeclareMathSymbol{\tvarRho}{\mathalpha}{tgreeki}{"50}
 \DeclareMathSymbol{\tvarSigma}{\mathalpha}{tgreeki}{"06}
 \DeclareMathSymbol{\tvarTau}{\mathalpha}{tgreeki}{"54}
 \DeclareMathSymbol{\tvarUpsilon}{\mathalpha}{tgreeki}{"07}
 \DeclareMathSymbol{\tvarPhi}{\mathalpha}{tgreeki}{"08}
 \DeclareMathSymbol{\tvarChi}{\mathalpha}{tgreeki}{"58}
 \DeclareMathSymbol{\tvarPsi}{\mathalpha}{tgreeki}{"09}
 \DeclareMathSymbol{\tvarOmega}{\mathalpha}{tgreeki}{"0A} 
}

\newcommand{\echgreek}{%
\egreeki
\let\alpha\ealpha
\let\beta\ebeta
\let\gamma\egamma
\let\delta\edelta
\let\epsilon\eepsilon
\let\zeta\ezeta
\let\eta\eeta
\let\theta\etheta
\let\iota\eiota
\let\kappa\ekappa
\let\lambda\elambda
\let\mu\emu
\let\nu\enu
\let\xi\exi
\let\omicron\eomicron
\let\pi\epi
\let\rho\erho
\let\sigma\esigma
\let\tau\etau
\let\upsilon\eupsilon
\let\phi\ephi
\let\chi\echi
\let\psi\epsi
\let\omega\eomega
\let\varepsilon\evarepsilon
\let\varphi\evarphi
\let\varpi\evarpi
\let\vartheta\evartheta
}

\newcommand{\echgreekup}{
\egreekr
\let\partialup\epartialup
\let\alphaup\ealphaup
\let\betaup\ebetaup
\let\gammaup\egammaup
\let\deltaup\edeltaup
\let\epsilonup\eepsilonup
\let\zetaup\ezetaup
\let\etaup\eetaup
\let\thetaup\ethetaup
\let\iotaup\eiotaup
\let\kappaup\ekappaup
\let\lambdaup\elambdaup
\let\muup\emuup
\let\nuup\enuup
\let\xiup\exiup
\let\omicronup\eomicronup
\let\piup\epiup
\let\rhoup\erhoup
\let\sigmaup\esigmaup
\let\tauup\etauup
\let\upsilonup\eupsilonup
\let\phiup\ephiup
\let\chiup\echiup
\let\psiup\epsiup
\let\omegaup\eomegaup
\let\varepsilonup\evarepsilonup
\let\varphiup\evarphiup
\let\varpiup\evarpiup
\let\varsigmaup\evarsigmaup
\let\varrhoup\evarrhoup
\let\varkappaup\evarkappaup
\let\varthetaup\evarthetaup
}
\newcommand{\echvarGreek}{
\egreeki
\let\varAlpha\evarAlpha
\let\varBeta\evarBeta
\let\varGamma\evarGamma
\let\varDelta\evarDelta
\let\varEpsilon\evarEpsilon
\let\varZeta\evarZeta
\let\varEta\evarEta
\let\varTheta\evarTheta
\let\varIota\evarIota
\let\varKappa\evarKappa
\let\varLambda\evarLambda
\let\varMu\evarMu
\let\varNu\evarNu
\let\varXi\evarXi
\let\varOmicron\evarOmicron
\let\varPi\evarPi
\let\varRho\evarRho
\let\varSigma\evarSigma
\let\varTau\evarTau
\let\varUpsilon\evarUpsilon
\let\varPhi\evarPhi
\let\varChi\evarChi
\let\varPsi\evarPsi
\let\varOmega\evarOmega
\let\varAlpha\evarAlpha
\let\varBeta\evarBeta
\let\varGamma\evarGamma
\let\varDelta\evarDelta
\let\varEpsilon\evarEpsilon
\let\varZeta\evarZeta
\let\varEta\evarEta
\let\varTheta\evarTheta
\let\varIota\evarIota
\let\varKappa\evarKappa
\let\varLambda\evarLambda
\let\varMu\evarMu
\let\varNu\evarNu
\let\varXi\evarXi
\let\varOmicron\evarOmicron
\let\varPi\evarPi
\let\varRho\evarRho
\let\varSigma\evarSigma
\let\varTau\evarTau
\let\varUpsilon\evarUpsilon
\let\varPhi\evarPhi
\let\varChi\evarChi
\let\varPsi\evarPsi
\let\varOmega\evarOmega
}
\newcommand{\echGreek}{
\egreekr
\let\Alpha\eAlpha
\let\Beta\eBeta
\let\Gamma\eGamma
\let\Delta\eDelta
\let\Epsilon\eEpsilon
\let\Zeta\eZeta
\let\Eta\eEta
\let\Theta\eTheta
\let\Iota\eIota
\let\Kappa\eKappa
\let\Lambda\eLambda
\let\Mu\eMu
\let\Nu\eNu
\let\Xi\eXi
\let\Omicron\eOmicron
\let\Pi\ePi
\let\Rho\eRho
\let\Sigma\eSigma
\let\Tau\eTau
\let\Upsilon\eUpsilon
\let\Phi\ePhi
\let\Chi\eChi
\let\Psi\ePsi
\let\Omega\eOmega
\let\Alpha\eAlpha
\let\Beta\eBeta
\let\Gamma\eGamma
\let\Delta\eDelta
\let\Epsilon\eEpsilon
\let\Zeta\eZeta
\let\Eta\eEta
\let\Theta\eTheta
\let\Iota\eIota
\let\Kappa\eKappa
\let\Lambda\eLambda
\let\Mu\eMu
\let\Nu\eNu
\let\Xi\eXi
\let\Omicron\eOmicron
\let\Pi\ePi
\let\Rho\eRho
\let\Sigma\eSigma
\let\Tau\eTau
\let\Upsilon\eUpsilon
\let\Phi\ePhi
\let\Chi\eChi
\let\Psi\ePsi
\let\Omega\eOmega
}

\newcommand{\tchgreek}{%
\tgreeki
\let\alpha\talpha
\let\beta\tbeta
\let\gamma\tgamma
\let\delta\tdelta
\let\epsilon\tepsilon
\let\zeta\tzeta
\let\eta\teta
\let\theta\ttheta
\let\iota\tiota
\let\kappa\tkappa
\let\lambda\tlambda
\let\mu\tmu
\let\nu\tnu
\let\xi\txi
\let\omicron\tomicron
\let\pi\tpi
\let\rho\trho
\let\sigma\tsigma
\let\tau\ttau
\let\upsilon\tupsilon
\let\phi\tphi
\let\chi\tchi
\let\psi\tpsi
\let\omega\tomega
\let\varepsilon\tvarepsilon
\let\varphi\tvarphi
\let\varpi\tvarpi
\let\vartheta\tvartheta
}
\newcommand{\tchgreekup}{
\tgreekr
\let\alphaup\talphaup
\let\betaup\tbetaup
\let\gammaup\tgammaup
\let\deltaup\tdeltaup
\let\epsilonup\tepsilonup
\let\zetaup\tzetaup
\let\etaup\tetaup
\let\thetaup\tthetaup
\let\iotaup\tiotaup
\let\kappaup\tkappaup
\let\lambdaup\tlambdaup
\let\muup\tmuup
\let\nuup\tnuup
\let\xiup\txiup
\let\omicronup\tomicronup
\let\piup\tpiup
\let\rhoup\trhoup
\let\sigmaup\tsigmaup
\let\tauup\ttauup
\let\upsilonup\tupsilonup
\let\phiup\tphiup
\let\chiup\tchiup
\let\psiup\tpsiup
\let\omegaup\tomegaup
\let\varepsilonup\tvarepsilonup
\let\varphiup\tvarphiup
\let\varpiup\tvarpiup
\let\varsigmaup\tvarsigmaup
\let\varrhoup\tvarrhoup
\let\varkappaup\tvarkappaup
\let\varthetaup\tvarthetaup
}
\newcommand{\tchvarGreek}{
\tgreeki
\let\varAlpha\tvarAlpha
\let\varBeta\tvarBeta
\let\varGamma\tvarGamma
\let\varDelta\tvarDelta
\let\varEpsilon\tvarEpsilon
\let\varZeta\tvarZeta
\let\varEta\tvarEta
\let\varTheta\tvarTheta
\let\varIota\tvarIota
\let\varKappa\tvarKappa
\let\varLambda\tvarLambda
\let\varMu\tvarMu
\let\varNu\tvarNu
\let\varXi\tvarXi
\let\varOmicron\tvarOmicron
\let\varPi\tvarPi
\let\varRho\tvarRho
\let\varSigma\tvarSigma
\let\varTau\tvarTau
\let\varUpsilon\tvarUpsilon
\let\varPhi\tvarPhi
\let\varChi\tvarChi
\let\varPsi\tvarPsi
\let\varOmega\tvarOmega
\let\varAlpha\tvarAlpha
\let\varBeta\tvarBeta
\let\varGamma\tvarGamma
\let\varDelta\tvarDelta
\let\varEpsilon\tvarEpsilon
\let\varZeta\tvarZeta
\let\varEta\tvarEta
\let\varTheta\tvarTheta
\let\varIota\tvarIota
\let\varKappa\tvarKappa
\let\varLambda\tvarLambda
\let\varMu\tvarMu
\let\varNu\tvarNu
\let\varXi\tvarXi
\let\varOmicron\tvarOmicron
\let\varPi\tvarPi
\let\varRho\tvarRho
\let\varSigma\tvarSigma
\let\varTau\tvarTau
\let\varUpsilon\tvarUpsilon
\let\varPhi\tvarPhi
\let\varChi\tvarChi
\let\varPsi\tvarPsi
\let\varOmega\tvarOmega
}
\newcommand{\tchGreek}{
\tgreekr
\let\Alpha\tAlpha
\let\Beta\tBeta
\let\Gamma\tGamma
\let\Delta\tDelta
\let\Epsilon\tEpsilon
\let\Zeta\tZeta
\let\Eta\tEta
\let\Theta\tTheta
\let\Iota\tIota
\let\Kappa\tKappa
\let\Lambda\tLambda
\let\Mu\tMu
\let\Nu\tNu
\let\Xi\tXi
\let\Omicron\tOmicron
\let\Pi\tPi
\let\Rho\tRho
\let\Sigma\tSigma
\let\Tau\tTau
\let\Upsilon\tUpsilon
\let\Phi\tPhi
\let\Chi\tChi
\let\Psi\tPsi
\let\Omega\tOmega
\let\Alpha\tAlpha
\let\Beta\tBeta
\let\Gamma\tGamma
\let\Delta\tDelta
\let\Epsilon\tEpsilon
\let\Zeta\tZeta
\let\Eta\tEta
\let\Theta\tTheta
\let\Iota\tIota
\let\Kappa\tKappa
\let\Lambda\tLambda
\let\Mu\tMu
\let\Nu\tNu
\let\Xi\tXi
\let\Omicron\tOmicron
\let\Pi\tPi
\let\Rho\tRho
\let\Sigma\tSigma
\let\Tau\tTau
\let\Upsilon\tUpsilon
\let\Phi\tPhi
\let\Chi\tChi
\let\Psi\tPsi
\let\Omega\tOmega
}

\newcommand{\pchgreek}{%
\pgreeki
\let\alpha\palpha
\let\beta\pbeta
\let\gamma\pgamma
\let\delta\pdelta
\let\epsilon\pepsilon
\let\zeta\pzeta
\let\eta\peta
\let\theta\ptheta
\let\iota\piota
\let\kappa\pkappa
\let\lambda\plambda
\let\mu\pmu
\let\nu\pnu
\let\xi\pxi
\let\omicron\pomicron
\let\pi\ppi
\let\rho\prho
\let\sigma\psigma
\let\tau\ptau
\let\upsilon\pupsilon
\let\phi\pphi
\let\chi\pchi
\let\p\ppsi
\let\omega\pomega
\let\varepsilon\pvarepsilon
\let\varphi\pvarphi
\let\varpi\pvarpi
\let\vartheta\pvartheta
}
\newcommand{\pchgreekup}{
\pgreekr
\let\alphaup\palphaup
\let\betaup\pbetaup
\let\gammaup\pgammaup
\let\deltaup\pdeltaup
\let\epsilonup\pepsilonup
\let\zetaup\pzetaup
\let\etaup\petaup
\let\thetaup\pthetaup
\let\iotaup\piotaup
\let\kappaup\pkappaup
\let\lambdaup\plambdaup
\let\muup\pmuup
\let\nuup\pnuup
\let\xiup\pxiup
\let\omicronup\pomicronup
\let\piup\ppiup
\let\rhoup\prhoup
\let\sigmaup\psigmaup
\let\tauup\ptauup
\let\upsilonup\pupsilonup
\let\phiup\pphiup
\let\chiup\pchiup
\let\psiup\ppsiup
\let\omegaup\pomegaup
\let\varepsilonup\pvarepsilonup
\let\varphiup\pvarphiup
\let\varpiup\pvarpiup
\let\varsigmaup\pvarsigmaup
\let\varrhoup\pvarrhoup
\let\varkappaup\pvarkappaup
\let\varthetaup\pvarthetaup
}
\newcommand{\pchvarGreek}{
\pgreeki
\let\varAlpha\pvarAlpha
\let\varBeta\pvarBeta
\let\varGamma\pvarGamma
\let\varDelta\pvarDelta
\let\varEpsilon\pvarEpsilon
\let\varZeta\pvarZeta
\let\varEta\pvarEta
\let\varTheta\pvarTheta
\let\varIota\pvarIota
\let\varKappa\pvarKappa
\let\varLambda\pvarLambda
\let\varMu\pvarMu
\let\varNu\pvarNu
\let\varXi\pvarXi
\let\varOmicron\pvarOmicron
\let\varPi\pvarPi
\let\varRho\pvarRho
\let\varSigma\pvarSigma
\let\varTau\pvarTau
\let\varUpsilon\pvarUpsilon
\let\varPhi\pvarPhi
\let\varChi\pvarChi
\let\varPsi\pvarPsi
\let\varOmega\pvarOmega
\let\varAlpha\pvarAlpha
\let\varBeta\pvarBeta
\let\varGamma\pvarGamma
\let\varDelta\pvarDelta
\let\varEpsilon\pvarEpsilon
\let\varZeta\pvarZeta
\let\varEta\pvarEta
\let\varTheta\pvarTheta
\let\varIota\pvarIota
\let\varKappa\pvarKappa
\let\varLambda\pvarLambda
\let\varMu\pvarMu
\let\varNu\pvarNu
\let\varXi\pvarXi
\let\varOmicron\pvarOmicron
\let\varPi\pvarPi
\let\varRho\pvarRho
\let\varSigma\pvarSigma
\let\varTau\pvarTau
\let\varUpsilon\pvarUpsilon
\let\varPhi\pvarPhi
\let\varChi\pvarChi
\let\varPsi\pvarPsi
\let\varOmega\pvarOmega
}
\newcommand{\pchGreek}{
\pgreekr
\let\Alpha\pAlpha
\let\Beta\pBeta
\let\Gamma\pGamma
\let\Delta\pDelta
\let\Epsilon\pEpsilon
\let\Zeta\pZeta
\let\Eta\pEta
\let\Theta\pTheta
\let\Iota\pIota
\let\Kappa\pKappa
\let\Lambda\pLambda
\let\Mu\pMu
\let\Nu\pNu
\let\Xi\pXi
\let\Omicron\pOmicron
\let\Pi\pPi
\let\Rho\pRho
\let\Sigma\pSigma
\let\Tau\pTau
\let\Upsilon\pUpsilon
\let\Phi\pPhi
\let\Chi\pChi
\let\Psi\pPsi
\let\Omega\pOmega
\let\Alpha\pAlpha
\let\Beta\pBeta
\let\Gamma\pGamma
\let\Delta\pDelta
\let\Epsilon\pEpsilon
\let\Zeta\pZeta
\let\Eta\pEta
\let\Theta\pTheta
\let\Iota\pIota
\let\Kappa\pKappa
\let\Lambda\pLambda
\let\Mu\pMu
\let\Nu\pNu
\let\Xi\pXi
\let\Omicron\pOmicron
\let\Pi\pPi
\let\Rho\pRho
\let\Sigma\pSigma
\let\Tau\pTau
\let\Upsilon\pUpsilon
\let\Phi\pPhi
\let\Chi\pChi
\let\Psi\pPsi
\let\Omega\pOmega
}

\egreekr\egreeki%\pgreekr\pgreeki\tgreekr\tgreeki
\usepackage{bm}
\providecommand{\piup}{\epiup}%constant pi
\providecommand{\deltaup}{\edeltaup}%constant pi
\providecommand{\zetaup}{\ezetaup}%constant pi
\providecommand{\varepsilonup}{\evarepsilonup}%constant pi
\providecommand{\nequiv}{\not\equiv}

\newcommand{\de}{\partialup}%partial diff
\newcommand{\e}{\mathrm{e}}%Neper
\newcommand{\di}{\mathrm{d}}%differential

\DeclareMathOperator{\tr}{tr}%trace
\newcommand{\incr}{\Delta}%finite increment
\newcommand{\defin}{\stackrel{_\text{def}}{=}}

\newcommand{\lmatimplies}{\Rightarrow}%entails
\newcommand{\lmatiff}{\Leftrightarrow}%equivalent
\newcommand{\corr}{\mathrel{\hat{=}}}%corresponds to
\renewcommand{\le}{\leqslant}%less or equal
\renewcommand{\ge}{\geqslant}%greater or equal
\DeclareMathDelimiter{\lclose}{\mathopen}{operators}{"5B}{largesymbols}{"02}
\DeclareMathDelimiter{\rclose}{\mathclose}{operators}{"5D}{largesymbols}{"03}
\DeclareMathDelimiter{\lopen}{\mathopen}{operators}{"5D}{largesymbols}{"03}
\DeclareMathDelimiter{\ropen}{\mathclose}{operators}{"5B}{largesymbols}{"02}

\newcommand{\ket}[1]{\lvert#1\rangle}
\newcommand{\braket}[2]{\langle#1\mid#2\rangle}
\newcommand{\ketbra}[2]{\lvert#1\rangle\langle#2\rvert}
\newcommand{\set}[1]{\{#1\}}
\newcommand{\sect}{\S}% Sect.~
\newcommand{\sects}{\S\S}% Sect.~
\theoremstyle{definition}

\newcommand{\etc}{{etc.}}
\newcommand{\ie}{{i.e.}}
\newcommand{\Eg}{{E.g.}}
\newcommand{\eg}{{e.g.}}
\newcommand{\viz}{{viz.}}
\newcommand{\cf}{{cf.}}

\newcommand{\etal}{{et al.}}
\newcommand{\nbd}{\nobreakdash}%
\newcommand{\bd}{\hspace{0pt}}%
\newcommand{\povm}{positive-\bd operator-\bd valued measure}%POVM
\newcommand{\POVM}{POVM}%POVM

\newcommand{\cpm}{completely positive map}
\newcommand{\CPM}{CPM}

\newcommand{\avogn}{N_{\textrm{A}}}
\newcommand{\zT}{T}%temp
\newcommand{\zN}{n}%num mol
\newcommand{\zNp}{N}%num part
\newcommand{\zk}{R}%gas const
\newcommand{\zVf}{V_\text{f}}%num part
\newcommand{\zVi}{V_\text{i}}%num part
\newcommand{\zham}{\bm{H}}%hamilt
\newcommand{\zrho}{\bm{\rho}}%generic rho
\newcommand{\zphi}{\bm{\phi}}%phi st
\newcommand{\zpsi}{\bm{\psi}}%psi st
\newcommand{\zzp}{z^+}%psi st
\newcommand{\zzm}{z^-}%psi st
\newcommand{\zzpm}{z^\pm}%psi st
\newcommand{\zxp}{x^+}%psi st
\newcommand{\zxm}{x^-}%psi st
\newcommand{\zxpm}{x^\pm}%psi st
\newcommand{\zpa}{\bm{A}_1}%psi st
\newcommand{\zpb}{\bm{A}_2}%psi st
\newcommand{\zpm}{\bm{z}^\pm}%psi st
\newcommand{\zz}{\bm{z}^+}%psi st
\newcommand{\zzb}{\bm{z}^-}%psi st
\newcommand{\zx}{\bm{x}^+}%psi st
\newcommand{\zzk}{\ket{\zzp}}%psi st ket
\newcommand{\zxk}{\ket{\zxp}}%psi st ket
\newcommand{\zmix}{\bm{\lambda}}%rho st
\newcommand{\zab}{\bm{\alpha}^{\pm}}%psi st
\newcommand{\za}{\bm{\alpha}^{+}}%psi st
\newcommand{\zb}{\bm{\alpha}^{-}}%psi st
\newcommand{\zabp}{\alpha^\pm}%psi st
\newcommand{\zap}{\alpha^+}%psi st
\newcommand{\zbp}{\alpha^-}%psi st
\newcommand{\zabk}{\ket{\alpha^\pm}}%psi st
\newcommand{\primeds}{\tilde}%[1]{{#1}_{+}}%
\newcommand{\doubleds}{\grave}%[1]{{#1}_{-}}%
\newcommand{\zzup}{\primeds{z}^+}%psi st
\newcommand{\zzupm}{\primeds{z}^\pm}%psi st
\newcommand{\zzvp}{\doubleds{z}^+}%psi st
\newcommand{\zzvpm}{\doubleds{z}^\pm}%psi st
\newcommand{\zxvp}{\doubleds{x}^+}%psi st
\newcommand{\zxvpm}{\doubleds{x}^\pm}%psi st
\newcommand{\zxup}{\primeds{x}^+}%psi st
\newcommand{\zxupm}{\primeds{x}^\pm}%psi st
\newcommand{\zzum}{\primeds{z}^-}%psi st
\newcommand{\zzvm}{\doubleds{z}^-}%psi st
\newcommand{\zxvm}{\doubleds{x}^-}%psi st
\newcommand{\zxum}{\primeds{x}^-}%psi st
\newcommand{\zzu}{\primeds{\bm{z}}^{+}}%psi st
\newcommand{\zxv}{\doubleds{\bm{x}}^{+}}%psi st
\newcommand{\zmi}{\bm{\tau}}%psi st
\newcommand{\zwbB}{\primeds{\bm{\alpha}}^{\pm}}%psi st
\newcommand{\zwbBk}{\primeds{\alpha}^\pm}%psi st
\newcommand{\zwbBpk}{\primeds{\alpha}^+}%psi st
\newcommand{\zwbBmk}{\primeds{\alpha}^-}%psi st
\newcommand{\zwcC}{\doubleds{\bm{\alpha}}^{\pm}}%psi st
\newcommand{\zwcCk}{\doubleds{\alpha}^\pm}%psi st
\newcommand{\zwcCpk}{\doubleds{\alpha}^+}%psi st
\newcommand{\zwcCmk}{\doubleds{\alpha}^-}%psi st
\newcommand{\zwEF}{\bm{E}^{\pm}}%psi st
\newcommand{\zaup}{\zwbBpk}%psi st
\newcommand{\zavp}{\zwcCpk}%psi st
\newcommand{\zbup}{\zwbBmk}%psi st
\newcommand{\zbvp}{\zwcCmk}%psi st
\newcommand{\xxx}{\bm{\rho}}%psi st

\newcommand{\zAa}{{}^a\text{Ar}}%argon 1
\newcommand{\zAb}{{}^b\text{Ar}}%argon 2

\newcommand{\diaphragm}{membrane}
\newcommand{\diaphragms}{membranes}

\begin{document}
\hyphenation{
im-pli-cans
im-pli-cate
dis-tin-guish-abil-i-ty
prot-a-sis
apod-o-sis
}
\bibliographystyle{apsrevmana}

\author{\firstname{Piero G. Luca}\hspace{1.7ex}\surname{Mana}}

\email{mana@imit.kth.se}

\affiliation{Laboratory of Quantum Electronics and Quantum
Optics (QEO), Department of Microelectronics and
Information Technology (IMIT), Royal Institute of
Technology (KTH), Isafjordsgatan 22, SE-164\,40 Stockholm,
Sweden}

%\datesymd

\title{Distinguishability of non-orthogonal density
matrices does not imply violations of the second law%:\\ a
%physical preparation is not a density matrix
}

\thanks{Dedicated to the memory of Asher Peres}

\date{30 January 2005}

\begin{abstract}
The hypothetical possibility of distinguishing
preparations described by non-\bd orthogonal density matrices
does not necessarily imply a violation of the second law
of thermodynamics, as was instead stated by von~Neumann.
On the other hand, such a possibility would surely mean
that the particular density-matrix space (and related
Hilbert space) adopted would not be adequate to describe
the hypothetical new experimental facts. These points are
shown by making clear the distinction between physical
preparations and the density matrices which represent
them, and then comparing a ``quantum'' thermodynamic
analysis given by Peres with a ``classical'' one given by
Jaynes.
\end{abstract}

\maketitle

\section{Von~Neumann (and Peres) on orthogonality 
and the second law}

In \sect V.2 of von~Neumann's
\emph{\langgerman{{Mathematische} {Grundlagen} der
{Quantenmechanik}}}~\citep{vonneumann1932c} we find the
following two propositions
(p.~197):\footnote{``\langgerman{zwei Zust\"ande $\zphi$,
$\zpsi$ [\ldots]\ durch semipermeable W\"ande bestimmt
getrennt werden k\"onnen, wenn sie orthogonal sind}'', and
``\langgerman{sind $\zphi$, $\zpsi$ nicht orthogonal, so
widerspricht die Annahme einer solchen semipermeablen Wand
dem zweiten Hauptsatz''}.}
\begin{equation}\label{eq:first}
\parbox{.8\columnwidth}{two states $\zphi$, $\zpsi$
[\ldots]\ can certainly be separated by a semi-permeable
\diaphragm\ if they are orthogonal;}
\end{equation}
and the converse
\begin{equation}\label{eq:second}
\parbox{.8\columnwidth}{{if} $\zphi$, $\zpsi$ are not
orthogonal, then the assumption of such a semi-permeable
\diaphragm\ contradicts the second law [of
thermodynamics].}
\end{equation}
The demonstrations of these two statements given by
von~Neumann are based on the same ``thermodynamic
considerations'' by which he derives his entropy formula.
Peres, in his insightful and lucid book~\citep{peres1995},
also gave a demonstration to show that if we were able to
produce two semi-permeable \diaphragms\ which
unambiguously distinguish non-\bd orthogonal ``states'',
we could violate the second law of thermodynamics.
%this statement has been mentioned by
%Maruyama \etal~\citep{maruyamaetal2004})

The purpose of this paper is first to rephrase the two
statements above, substituting the ambiguous term
``state'' with other terms which make clear the
distinction between a physical phenomenon and its
mathematical
description~\citep{ekstein1967,ekstein1969,giles1968,giles1970,parketal1976,bandetal1976,giles1979,foulisetal1972a,randalletal1973b,foulisetal1978,peres1984,peres1995},
and then to show that von~Neumann's (and Peres') second
statement, once rephrased, is not necessarily true. This
will be done by comparing Peres'
demonstration~\citep{peres1995} with a lucid, simple, and
probably less known analysis, given by
Jaynes~\citep{jaynes1992}, of a seeming violation of the
second law of thermodynamics apparently due to
\emph{classical} (\ie\ non-\bd quantum)
``indistinguishability'' issues.
%% The importance of the
%% distinction between physical phenomenon and mathematical
%% description will then be reaffirmed and discussed in the
%% concluding remarks.

The paper is not directly concerned with questions about
the relation between thermodynamics and statistical
mechanics, nor to questions about ``classical'' or
``quantum'' entropy formulae. Indeed, \emph{no particular
entropy formula will be used}, but only the assumption
that in a closed thermodynamic cycle the entropy change
vanishes.

\section{Exegesis of von~Neumann's statements\label{sec:rephrase}}

Let me now rephrase von~Neumann's statements above, at the
cost of making some semantic violence to them, and in
particular let me eliminate the too-many-faced term
``state'' in favour of the two distinct terms
\emph{(physical)
preparation}~\citep{ekstein1967,ekstein1969,giles1968,giles1970,giles1979,foulisetal1972a,randalletal1973b,foulisetal1978,parketal1976,bandetal1976,peres1984,peres1995}
(\cf\ also~\citep{ballentine1970,hardy2001,mana2004b}), which has a clear
physical, experimental meaning, and \emph{statistical
matrix}, which denotes a mathematical object instead:
`statistical matrix' is the old term for what has been
called `density matrix' apparently since
Wigner~\citep{wigner1932} (\cf\ Fano~\citep{fano1957});
since the matrices we consider here concern statistics but
no probability densities, I prefer to use the old term
hereafter.

Let me also substitute, but only for the time being, the
thermodynamic concept of a semi-permeable \diaphragm\ with
the more general idea of an \emph{observation
procedure}~\citep{ekstein1967,ekstein1969,giles1968,giles1970,giles1979,foulisetal1972a,randalletal1973b,foulisetal1978,peres1984,peres1995}.
Von~Neumann spoke of \diaphragms\ only because he needed a
conceptual device that could endow with thermodynamic
consequences the ability to \emph{distinguish} (and thus
separate) physical preparations. We shall reintroduce and
discuss this device in the next section, but for the
moment let us be more general.

\medskip

We can express von~Neumann's first
statement~\eqref{eq:first} as follows:
%{\samepage
\begin{equation}\label{eq:firstbis}
\parbox{.8\columnwidth}{if there is an observation
procedure by which we can distinguish two physical
preparations with certainty, then we mathematically
represent the latter by orthogonal statistical
matrices; and vice versa, by non-\bd orthogonal statistical
matrices if we do not know of any such observation
procedure.}\tag{\ref{eq:first}'}
\end{equation}
%}
(Two statistical matrices $\zphi$ and $\zpsi$ are said to
be orthogonal if and only if $\tr(\zphi \zpsi) = 0$.) This
re-statement makes clear the difference between the
physical phenomenon and the mathematical objects used to
describe, or represent, it: we see that von~Neumann's
original proposition, which looked mathematical in nature,
disguises in fact a statement about physics methodology.
Note also that from a logical point of view\footnote{In
the following I use the two logical symbols
`$\lmatimplies$' (`implies', `if \ldots\ then') for
logical implication, and `$\land$' (`and') for logical
conjunction~\citep{copietal1953}.} I have completed
von~Neumann's original proposition
\begin{equation*}
\text{orthogonality}\lmatimplies\text{distinguishability}
\end{equation*}
into the two 
\begin{align*}
\text{distinguishability}
&\lmatimplies
\text{orthogonality},\\
\text{indistinguishability}
&\lmatimplies
\text{non-orthogonality},\\
\intertext{or equivalently}
\text{distinguishability}
&\lmatiff
\text{orthogonality}.
\end{align*}
%% \begin{equation*}
%% \text{\begin{tabular}{c}
%% {distinguishability}\\ \emph{(physical fact)}
%% \end{tabular}}
%% \;\lmatimplies\;
%% \text{\begin{tabular}{c}
%% {orthogonality}\\ \emph{(mathematical description)}
%% \end{tabular}}
%% \end{equation*}
%% have exchanged the
%% antecedent (also called implicans or protasis) with the
%% consequent (also called implicate or apodosis), \ie, I

(The reader should pay attention not to confuse the
\emph{``causal''} connexion, in a loose sense, with the
\emph{logical} connexion which exist between
distinguishability and orthogonality. The causal connexion
goes only in one direction: a physicist uses orthogonal
matrices \emph{because} the corresponding preparations are
distinguishable, but two preparations do not become
suddenly distinguishable just because she has represented
them on paper by two orthogonal matrices. The logical
connexion goes instead both ways: if a physicist can
distinguish two preparations, we can deduce that she will
represent them by orthogonal matrices; and if we see that
a physicist represents two preparations by orthogonal
matrices, we can deduce that she knows of an observation
procedure that can distinguish those preparations. The
present author confused these two kind of connexions
himself in the first drafts of this paper. This kind of
confusion between physics and probability is called ``the
mind-projection fallacy'' by
Jaynes~\citep{jaynes1989,jaynes1990,jaynes2003}.)

\medskip

With regard to von~Neumann's (and Peres') second
proposition \eqref{eq:second}, it can be rephrased as
follows:% in the same spirit
\begin{equation}\label{eq:secondbis}
\parbox{.8\columnwidth}{if we could distinguish, by means of some observation
procedure, two physical preparations represented by
non-\bd orthogonal statistical matrices, then we could violate
the second law of thermodynamics.}\tag{\ref{eq:second}'}
\end{equation}
This proposition makes as well a clear distinction between
physical phenomenon and mathematical description; however,
\emph{it is logically intrinsically vain}. Let us see why
from a logical point of view first. The antecedent of the
proposition\footnote{In a proposition of the form
`$A\lmatimplies B$', $A$ is called the \emph{antecedent}
(or \emph{implicans}, or \emph{protasis}) and $B$ the
\emph{consequent} (or \emph{implicate}, or
\emph{apodosis})~\citep{copietal1953}.} formally is
`distinguishability${}\land{}$non-\bd orthogonality', but
this is \emph{false} in view of the previous
statement~\eqref{eq:firstbis}, namely
`distinguishability${}\lmatimplies{}$orthogonality':
remember that if `$A\lmatimplies B$' holds, then we
\emph{cannot} have both $A$ true and $B$
false~\citep{copietal1953}.  But we know that from a false
antecedent one can idly deduce any proposition
whatever~\citep{copietal1953}, hence the
statements~\eqref{eq:second} and~\eqref{eq:secondbis} are
devoid of real content. Speaking on a less abstractly
logical level, the point is that if we can distinguish two
preparations represented by non-\bd orthogonal statistical
matrices, then we are evidently no longer following our
prescription~\eqref{eq:firstbis} to mathematically
represent preparation distinguishability by means of
matrix orthogonality; we are inconsistent. Any
consequences we derive, like \eg\ violations of the second
law, are likely to be only artifacts of our inconsistency
rather than real phenomena. We must thus amend the
particular statistical matrices or the whole
statistical-matrix space used, since they have not been
adequately chosen to describe the physical phenomenon in
question or our experimental capabilities.

\medskip

I shall in fact show explicitly that the violation of the
second law of statements~\eqref{eq:second}
and~\eqref{eq:secondbis} is only an artifact, by analysing
Peres' demonstration~\citep{peres1995} and comparing it
with Jaynes' already mentioned analysis~\citep{jaynes1992}
of an analogous seeming violation of the second law of
thermodynamics due to entirely classical
``indistinguishability'' issues.\footnote{Jaynes used his
analysis to show that the definition and the
quantification of the thermodynamic entropy depend on the
particular thermodynamic variables that \emph{define} the
thermodynamic system: a long known fact, which also
Grad~\citep{grad1952,grad1961,grad1967}, who had different
views on statistical mechanics than Jaynes', stressed.} In
order to do this, let us analyse the way in which
von~Neumann and Peres link quantum distinguishability and
non-\bd orthogonality with the second law of
thermodynamics.

\section{``Quantum'' ideal gases, semi-permeable \diaphragms, and thermodynamics\label{sec:matintro}}

In order to analyse a proposition which relates
distinguishability of quantum preparations with the second
law of thermodynamics, it is necessary to introduce a
thermodynamic body possessing ``quantum'' characteristics,
\ie, quantum degrees of freedom.

\medskip

Let us first recall that (``classical'') ideal gases are
defined as homogeneous, uniform thermodynamic bodies
characterisable by two thermodynamic variables: the volume
$V>0$ and the temperature $\zT>0$, and for which the
internal energy is a function of the variable $\zT$
alone;\footnote{See \eg\ the very fine little book by
Truesdell and Bharatha~\citep{truesdelletal1977}.} this
implies, via the first law of thermodynamics, that
in any isothermal process the work \emph{done by the gas},
$W$, is always equal to the heat \emph{absorbed by the
gas}, $Q$:
\begin{gather}
\label{eq:working}
W=\zN \zk \zT \ln (\zVf/\zVi) = Q \quad \text{(isothermal
processes)},
\end{gather}
where $\zVi$ and $\zVf$ are the volumes at the beginning
and end of the process,
%% which satisfy the constitutive equation $p V = \zN \zk
%% \zT$, 
$\zk$ is the molar gas constant, and $\zN$ is the
(constant) number of moles. This formula will be true
throughout the paper, as we shall only consider isothermal
processes.

One often considers samples of such ideal gases in a
chamber and inserts, moves, or removes impermeable or
semi-permeable
\diaphragms\footnote{Partington~\citep[\sect28]{partington1949}
informs us that these were first used in thermodynamics by
Gibbs~\citep{gibbs1876}.} at any position one pleases in
order to subject the samples, independently of each other,
to variations of \eg\ volume or
pressure.\footnote{Although the conclusions drawn from
such kind of reasonings are often valid, it must be said
that the mathematical description adopted is wanting
(Truesdell often denounced the fact that classical
thermodynamics has only very rarely been treated with the
conceptual respect and mathematical dignity which are
accorded to other theories like rational mechanics or
general relativity). The correct formalism to describe
insertions and removals of \diaphragms\ should involve
field quantities (\cf\ 
Buchdahl~\citep[\sects46,~75]{buchdahl1966} and see \eg\ 
Truesdell~\citep[lectures~5,~6 and related
appendices]{truesdell1969} or
references~\citep{truesdelletal1960,bowen1968,truesdell1968b,samohyl1987,samohyl1999}),
as indicated by the possibility of introducing as many
\diaphragm s as we wish and hence to control smaller and
smaller portions of the gas.}

\medskip

We must now face the question of how to introduce and
mathematically represent quantum degrees of freedom in
an ideal gas. Von~Neumann~\citep[\sect
V.2]{vonneumann1932c}
%% , in order to show the thermodynamic
%% validity of his entropy formula
%% ``$\tr(\zrho\ln\zrho)$'',
used a hybrid classical-quantum description,
microscopically modelling a ``quantum'' ideal gas as a
quantity $\zN$ of classical particles (for simplicity)
possessing an ``internal'' quantum degree of freedom
represented by a statistical matrix $\zrho$ living in an
appropriate statistical-matrix space; this space and the
statistical matrix are always assumed to be the same for
all the gas particles. He then treated two gas samples
described by different statistical matrices as gases of
somehow \emph{different chemical species}.%
%% the problem is to define exactly what
%% one means by ``different gases'' when the statistical
%% matrices related to the two gases 
%% are non-\bd orthogonal, as he remarks on
%% p.~196.
\footnote{The idea had been presented by Einstein eighteen
years earlier~\citep{einstein1914}, but it is important to
point out that for Einstein the ``internal quantum degree
of freedom'' was just a ``resonator'' capable of assuming
only discrete energies (\ie, it was not described by
statistical matrices, and non-\bd orthogonality issues were
unknown); thus his application was less open to problems
and critique than von~Neumann's.} This conceptual device
presents some problems, to be discussed more in detail
elsewhere~\citep{bjoerketal2004b};
%% (see \eg\ the problem with
%% isochoric unitary rotations in the next section,
%% footnote~\ref{fn:isochoric})
for example, the chemical species of a gas is not a
thermodynamic variable, and even less a continuous one:
chemical differences cannot change \emph{continuously} to
zero.\footnote{An observation made by
Partington~\citep[\sect II.28]{partington1949} in
reference to Larmor~\citep[p.~275]{larmor1897}.}
Intuitively, it would seem more appropriate to somehow
describe a quantum ideal gas by the variables $(V, \zT,
\zrho)$ instead, taking values on appropriate sets.

However, we shall follow von~Neumann and Peres instead and
speak of a `$\zphi$-gas', or a `$\zpsi$-gas', \etc, where
$\zphi$ or $\zpsi$ are the statistical matrices
describing its quantum degrees of freedom, just as if we
were speaking of gases of different chemical species (like
\eg\ `argon' and `helium').\footnote{Dieks and
van~Dijk~\citep{dieksetal1988} point rightly out that for
a $\zphi$-gas one should more correctly consider the total
statistical matrix $\bigotimes_{i=1}^{\zNp}\zphi$, where
$\zNp = \zN \avogn$ (with $\avogn$ Avogadro's constant) is
the total number of particles.\label{fn:dieks}} The
thermodynamic variables are only $(V, \zT)$ for each such
gas.

\medskip

Two samples of these quantum ideal gases can be (more or
less effectively) separated by semi-permeable \diaphragms,
analogous to those used with chemically different gases.
The microscopic
idea~\citep[p.~196]{vonneumann1932c}\citep[p.~271]{peres1995}
is, paraphrasing von~Neumann, to construct many windows in
the \diaphragm, each of which is made as follows: each
particle of the gases is detained there and an observation
is performed on its quantum degrees of freedom. Depending
on the observation result, the particle penetrates the
window or is reflected, with unchanged momentum. This
implies that the number of particles and hence the
pressures or volumes of the gases on the two sides of the
semi-permeable \diaphragm\ will vary, and may set the
membrane in motion, producing work (\eg\ by lifting a
weight which loads the membrane). In more general terms,
this \diaphragm\ is a device which performs a physical
operation, described by a given \povm\ (\POVM) and \cpm\ 
(\CPM)~\citep{davies1983,kraus1983,peres2000a,peres2000b},
on the quantum degrees of freedom of the gases' particles,
and depending on the result it acts on their translational
degrees of freedom, \eg\ separating them spatially. There
arises thus a kind of mutual dependence between the
quantum degrees of freedom and the thermodynamic
parameters (like the volume $V$) of a quantum ideal gas.

\medskip

\begin{figure}[t]
\setlength{\unitlength}{0.0047\columnwidth}
\begin{picture}(140,50)(0,0)
\put(0,0){\framebox(55,40){\scriptsize$\begin{aligned}&0.5\,\ketbra{\zzp}{\zzp}+\\&0.5\,\ketbra{\zzm}{\zzm}\end{aligned}$}}
%\put(0,0){\framebox(55,20){\scriptsize$\ketbra{\zxp}{\zxp}$}}
\put(0,41){\makebox(0,0)[bl]{\scriptsize(a)}}

\put(57,20){\vector(1,0){18.5}}
\put(66.25,21){\makebox(0,0)[b]{\scriptsize$Q<0$}}
%\put(66.25,19){\makebox(0,0)[t]{\scriptsize?}}

\put(77.5,20){\framebox(55,20){\scriptsize$\ketbra{\zzp}{\zzp}$}}
\put(77.5,0){\framebox(55,20){\scriptsize$\ketbra{\zzm}{\zzm}$}}
\put(77.5,41){\makebox(0,0)[bl]{\scriptsize(b)}}
\end{picture}
\caption{Separation of completely distinguishable quantum gases}\label{fig:distin}
\end{figure}
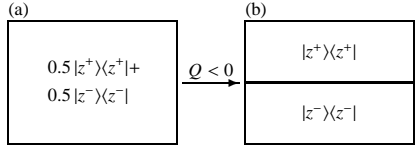

Let us make an example. Imagine a chamber having volume
$V$ and containing a mixture of a quantity $\zN/2$ of a
$\zz$-gas and $\zN/2$ of a $\zzb$-gas, \ie, of two quantum
ideal gases whose quantum degrees of freedom are
represented by the statistical matrices
\begin{align}
\zz &\defin \ketbra{\zzp}{\zzp} \corr
\Bigl(\begin{smallmatrix} 1&0\\0&0 
\end{smallmatrix}\Bigr),\label{eq:zplus}\\
\zzb &\defin \ketbra{\zzm}{\zzm} \corr
\Bigl(\begin{smallmatrix} 0&0\\0&1
\end{smallmatrix}\Bigr),
\end{align}
in the usual spin-1/2 notation (Fig.~\ref{fig:distin},
step a). These two statistical matrices represent quantum
preparations that can be distinguished with certainty,
hence there exists a semi-permeable \diaphragm, of the
kind described above, which is completely opaque to (the
particles of) the $\zzb$-gas and completely transparent to
(the particles of) the $\zz$-gas. Analogously, there
exists another semi-permeable \diaphragm\ with the
opposite properties, \viz\ it is completely opaque to (the
particles of) the $\zz$-gas and completely transparent to
(the particles of) the $\zzb$-gas. Such \diaphragm s
implement the two-element \POVM\ $\set{\zpm}$ with
outcomes given by the \CPM s $\set{\xxx \mapsto
\zpm\xxx\zpm/ \tr(\zpm\xxx\zpm)}$: acting on the $\zz$-gas
it yields\footnote{Due to the large number of gas
particles considered, the outcome probabilities are
numerically equal, within small fluctuations negligible in
the present work, to the average fraction of gas
correspondingly transmitted or reflected by the
\diaphragms.}
\begin{subequations}
\label{eq:actplus}
\begin{gather}
\begin{aligned}
\zz &\mapsto %\frac{\zz\zz\zz}{\tr(\zz\zz\zz)} = 
\zz&&\text{ with probability $\tr(\zz\zz\zz)=1$},\\
\zz &\mapsto %\frac{\zzb\zz\zzb}{\tr(\zzb\zz\zzb)} = 
\zzb&&\text{ with probability $\tr(\zzb\zz\zzb)=0$},
\end{aligned}\\
\intertext{while acting on the $\zzb$-gas it yields}
\begin{aligned}
\zzb &\mapsto %\frac{\zz\zzb\zz}{\tr(\zz\zzb\zz)} = 
\zz&&\text{ with probability $\tr(\zz\zzb\zz)=0$},\\
\zzb &\mapsto %\frac{\zzb\zzb\zzb}{\tr(\zzb\zzb\zzb)} = 
\zzb&&\text{ with probability $\tr(\zzb\zzb\zzb)=1$}
\end{aligned}
\end{gather}
\end{subequations}
(note that these are just von~Neumann projections), \ie,
the operation separates the two preparations which
certainty. The only difference between the \diaphragm s is
in which kind of gas (particle) they let through.

Now, imagine to insert these two \diaphragm s in the
chamber, very near to its top and bottom walls
respectively. Pushing them isothermally toward the middle
of the chamber, they will separate the two gases and in
the end we shall have the $\zz$-gas completely above the
first \diaphragm\ and the $\zzb$-gas completely under the
other in contact with the first (step b). In order to
move these \diaphragm s and achieve this separation
\emph{we} have to \emph{spend} an amount of work equal to
\begin{equation}\label{eq:worksepar}
- 2\times \frac{\zN}{2} \zk \zT \ln\frac{V/2}{V}
% =\zN \zk \zT \ln 2
\approx 0.693\, \zN \zk \zT,
\end{equation}
because each \diaphragm\ must overcome the pressure
exerted by the gas to which it is opaque. Since the
process is isothermal, the quantity above is also the
(positive) amount of heat \emph{released by the gases}.

The semi-permeable \diaphragm s can also be used to
realise the inverse process, \ie\ the mixing of two
initially separated $\zz$- and $\zzb$-gases. In this case
Eq.~\eqref{eq:worksepar} would be the amount of heat
\emph{absorbed by the gases} as well as the amount of work
\emph{performed} by them.

\medskip

However, the quantum degrees of freedom of two gases may
also be prepared in such a way that no observation
procedure can distinguish between them with certainty, and
hence there is no semi-permeable \diaphragm\ which can
separate them completely; this has of course consequences
for the amount of work that can be gained by using the
\diaphragm. Let us make another example. Imagine again the
initial situation above, but this time with a mixture of a
quantity $\zN/2$ of a $\zz$-gas and $\zN/2$ of an
$\zx$-gas, with
\begin{equation}
\zx \defin \ketbra{\zxp}{\zxp} \corr \tfrac{1}{2}
\Bigl(\begin{smallmatrix} 1&1\\1&1
\end{smallmatrix}\Bigr).\label{eq:xplus}
\end{equation}
The two statistical matrices $\zz$ and $\zx$ are non-\bd
orthogonal, $\tr(\zz\zx)\neq 0$, and correspond to quantum
preparations that cannot be distinguished with certainty,
and so there do not exist semi-permeable \diaphragm s
which are completely opaque to (the particles of) the one
gas and completely transparent to (the particles of) the
other gas. Mathematically this is reflected in the non-\bd
existence of some two-outcome \POVM\ $\set{\zpa, \zpb}$
with the properties
\begin{subequations}
\label{eq:actnondist}
\begin{gather}
\begin{aligned}
&\zz \text{ yields first outcome
with probability $\tr(\zpa\zz\zpa)=1$},\\
&\zz \text{ yields second outcome with probability
$\tr(\zpb\zz\zpb)=0$},
\end{aligned}\\
\intertext{and}
\begin{aligned}
&\zx \text{ yields first outcome
with probability $\tr(\zpa\zx\zpa)=0$},\\
&\zx \text{ yields second outcome with probability
$\tr(\zpb\zx\zpb)=1$}.
\end{aligned}
\end{gather}
\end{subequations}

It can be shown~\citep{vonneumann1932c,peres1995} that in
this case the operation which will require the maximum
amount of work is represented by the two-element \POVM\ 
$\set{\zab}$, where 
\begin{gather} 
\zab \defin
\ketbra{\zabp}{\zabp} \corr \frac{1}{4}\begin{pmatrix}
2\pm\sqrt{2}&\pm\sqrt{2}\\ \pm\sqrt{2}& 2\mp \sqrt{2}
\end{pmatrix},\\
\begin{split}\label{eq:newketalf}
\zabk &\defin \Bigl(2\pm \sqrt{2}\Bigr)^{-\frac{1}{2}}
\bigl(\ket{\zzpm} \pm \ket{\zxpm}\bigr), \\
&\equiv
\frac{1}{2}\Bigl[\pm\Bigl(2\pm\sqrt{2}\Bigr)^{\frac{1}{2}}
\ket{\zzp} + \Bigl(2\mp\sqrt{2}\Bigr)^{\frac{1}{2}}
\ket{\zzm}\Bigr],
\end{split}
\end{gather}
whose outcomes are given by the \CPM s (projections) $\xxx
\mapsto \zab\xxx\zab/ \tr(\zab\xxx\zab)$. Note that the
Hilbert-space vectors $\zabk$ are the eigenvectors of the
matrix $\zmix$, where
\begin{equation}
\label{eq:mixrho}
\zmix \defin \frac{1}{2}\zz % \ketbra{\zzp}{\zzp} 
+ \frac{1}{2} \zx %\ketbra{\zxp}{\zxp} 
= \frac{2 + \sqrt{2}}{4} \za
+ \frac{2 - \sqrt{2}}{4} \zb
\corr \frac{1}{4}\begin{pmatrix}
3&1\\1&1
\end{pmatrix},
\end{equation}
and they are orthogonal, so that $\tr(\za\zb)=0$.
The action of this \POVM\ on the statistical matrices of
our gases is given by
\begin{subequations}\label{eq:modifsepare}
\begin{gather}
\begin{aligned}
\zz &\mapsto %\frac{\za\zz\za}{\tr(\za\zz\za)} = 
\za&&\text{ with probability }
\begin{aligned}[t]
\tr(\za\zz\za) &=\bigl(2 + \sqrt{2}\bigr)/4\\
&\approx 0.854,
\end{aligned}
\\
\zz &\mapsto %\frac{\zzb\zz\zzb}{\tr(\zzb\zz\zzb)} = 
\zb&&\text{ with probability }
\begin{aligned}[t]
\tr(\zb\zz\zb)&=\bigl(2 - \sqrt{2}\bigr)/4\\
&\approx 0.146,
\end{aligned}
\end{aligned}\\
\intertext{and}
\begin{aligned}
\zx &\mapsto %\frac{\za\zx\za}{\tr(\zz\zzb\zz)} = 
\za&&\text{ with probability $\tr(\za\zx\za)\approx 0.854$},\\
\zx &\mapsto %\frac{\zzb\zzb\zzb}{\tr(\zzb\zzb\zzb)} = 
\zb&&\text{ with probability $\tr(\zb\zx\zb)\approx
0.146$}.
\end{aligned}
\end{gather}
\end{subequations}
The significance of the equations above is that the
half/half mixture of $\zz$- and $\zx$-gases can
equivalently be treated as a $0.854/0.146$ mixture of an
$\za$-gas and an $\zb$-gas, as is now shown.

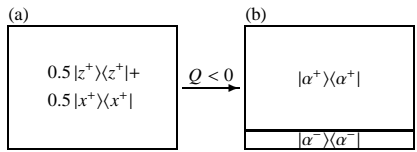
\begin{figure}[t]
\setlength{\unitlength}{0.0047\columnwidth}
\begin{picture}(140,50)(0,0)
\put(0,0){\framebox(55,40){\scriptsize$\begin{aligned}&0.5\,\ketbra{\zzp}{\zzp}+\\&0.5\,\ketbra{\zxp}{\zxp}\end{aligned}$}}
%\put(0,0){\framebox(55,20){\scriptsize$\ketbra{\zxp}{\zxp}$}}
\put(0,41){\makebox(0,0)[bl]{\scriptsize(a)}}

\put(57,20){\vector(1,0){18.5}}
\put(66.25,21){\makebox(0,0)[b]{\scriptsize$Q<0$}}
%\put(66.25,19){\makebox(0,0)[t]{\scriptsize?}}

\put(77.5,6){\framebox(55,34){\scriptsize$\ketbra{\zap}{\zap}$}}
\put(77.5,0){\framebox(55,6){\scriptsize$\ketbra{\zbp}{\zbp}$}}
\put(77.5,41){\makebox(0,0)[bl]{\scriptsize(b)}}
\end{picture}
\caption{Separation of partially distinguishable quantum gases}\label{fig:nondistin}
\end{figure}

We can construct two semi-permeable \diaphragm s which
implement the above \POVM\ and \CPM s in such a way that
they can be used to completely separate an $\za$-gas from
an $\zb$-gas, \ie, one \diaphragm\ is totally transparent
to the former and totally opaque to the latter, and vice
versa for the other \diaphragm. By inserting these
\diaphragm s in the chamber as in the preceding example,
they will partially separate \emph{and transform} our
$\zz$- and $\zx$-gases leaving in the end an $\za$-gas
above and an $\zb$-gas below the two (adjoined) \diaphragm
s (Fig.~\ref{fig:nondistin}). The gases will have same the
pressure but occupy unequal volumes because the ratio for
both $\zz$- and $\zx$-gases to be transformed into an
$\za$-gas and an $\zb$-gas is approximately $0.854/0.146$,
as seen from Eqs.~\eqref{eq:modifsepare}; this will also
be the ratio of the final volumes. The total amount of
work necessary to perform this separation is
\begin{multline}\label{eq:worknoondingsepar}
-0.146\, \zN \zk \zT \ln \frac{0.146 V}{V} -
0.854\, \zN \zk \zT \ln \frac{0.854 V}{V}
\approx{}\\ 0.416\, \zN \zk \zT,
\end{multline}
where the first term is for the upper \diaphragm\ 
(transparent to the $\za$-gas) and the second for the
lower one (transparent to the $\zb$-gas). This is also the
amount of heat \emph{released by the gases}, and we see
that it is less than the amount for the previous case,
Eq.~\eqref{eq:worksepar}. Of course, in the reverse
process, in which we mix two initially separated
$\zab$-gases in the same amounts, the \emph{gases} would
\emph{absorb} the same (positive) amount of heat.

%% \begin{gather}\label{eq:actnondist}
%% \begin{aligned}
%% \zz &\mapsto \text{(something)}%\frac{\zpa\zz\zpa}{\tr(\zpa\zz\zpa)} 
%% &&\text{with probability $\tr(\zpa\zz\zpa)=1$},\\
%% \zz &\mapsto 0%\frac{\zpb\zz\zpb}{\tr(\zpb\zz\zpb)} 
%% &&\text{with probability $\tr(\zpb\zz\zpb)=0$},
%% \end{aligned}\\
%% \intertext{and}
%% \begin{aligned}
%% \zx &\mapsto 0%\frac{\zpa\zx\zpa}{\tr(\zpa\zx\zpa)} 
%% &&\text{with probability $\tr(\zpa\zx\zpa)=0$},\\
%% \zx &\mapsto \text{(something else)}%\frac{\zpb\zx\zpb}{\tr(\zpb\zx\zpb)} 
%% &&\text{with probability $\tr(\zpb\zx\zpb)=1$}
%% \end{aligned}
%% \end{gather}

\medskip

We shall in a moment examine how Peres uses such
semi-permeable \diaphragm s to show that if we could
distinguish and separate two quantum-ideal-gas samples
characterised by non-\bd orthogonal statistical matrices,
then a violation of the second law would follow. Although
he explicitly adopts von~Neumann's entropy formula, his
demonstration really only uses the assumption that the
thermodynamic entropy depends on the values of the
parameters describing the phenomena in question at a
particular instant but not on their history (as instead is
the case in more general thermodynamic and thermomechanic
processes~\citep{truesdell1969}), so that in a cyclic
process, in which we start from and end in a situation
described by identical parameter values, the entropy
change is naught:
\begin{equation}
\label{eq:entropycycle}
\incr S = 0\quad\text{(cyclic process)}.
\end{equation}
\emph{We shall also make this same sole assumption}, and
since it does not require a specific formula for the
entropy (such as von~Neumann's formula), we shall
\emph{not} make use of any particular entropy expression.
%% However, I shall not touch issues related to the
%% (``classical'' or ``quantum'') entropy, \emph{and will not
%% use any particular entropy formula:} the only assumption
%% made in the following discussion is that the thermodynamic
%% entropy depends on the values of the parameters describing
%% the phenomena in question at a particular instant but not
%% on their history (as instead is the case in more general
%% thermodynamic and thermomechanic
%% processes~\citep{truesdell1969}), so that in a cyclic
%% process, in which we start from and end in a situation
%% described by identical parameter values, the entropy
%% change is naught.

Finally, let us recall that the second law of
thermodynamics for isothermal processes says that the
amount of heat $Q$ \emph{absorbed} by a thermodynamic
body, divided by the temperature $T$, is bounded above by
the change in entropy:\footnote{This is a special case of
the Clausius-Duhem
inequality~\citep[\sect258]{truesdelletal1960}\citep[lecture~2]{truesdell1969}\citep{silhavy1997},
which is more generally valid for inhomogeneous,
non-uniform bodies and for irreversible, non necessarily
isothermal processes: $\int_B r/T\, \di m + \int_{\de B}
q/T\, \di A \le \int_B \dot{s} \, \di m$, where $r$, $q$,
$s$, are respectively the massic heating supply, the
heating influx, the massic entropy of the body $B$ having
surface $\de B$, and the dot represents the substantial
derivative.}
\begin{equation}
\label{eq:seclaw}
Q/\zT\le \incr S\quad\text{(isothermal process)}.
\end{equation}
This assumes in our case for a cyclic process the
form\footnote{See also Serrin's~\citep{serrin1979} nice
analysis of the second law for cyclic processes.}
\begin{equation}
\label{eq:seclawcycl}
Q/\zT \le 0 = \incr S\quad
\text{(isothermal cyclic process)}.
\end{equation}

\section{Peres' demonstration\label{sec:peresdem}}

\begin{figure}[b]
\setlength{\unitlength}{0.0047\columnwidth}
\begin{picture}(210,109)(0,-60)
\put(0,20){\framebox(55,20){\scriptsize$\ketbra{\zzp}{\zzp}$}}
\put(0,0){\framebox(55,20){\scriptsize$\ketbra{\zxp}{\zxp}$}}
\put(0,41){\makebox(0,0)[bl]{\scriptsize(a)}}

\put(57,20){\vector(1,0){18.5}}
\put(66.25,21){\makebox(0,0)[b]{\scriptsize$Q'>0$}}
\put(66.25,19){\makebox(0,0)[t]{\scriptsize?}}

\put(77.5,0){\framebox(55,40){\scriptsize$\begin{aligned}&0.5\,\ketbra{\zzp}{\zzp}+\\&0.5\,\ketbra{\zxp}{\zxp}\end{aligned}$}}
\put(77.5,41){\makebox(0,0)[bl]{\scriptsize(b)}}

\put(134.5,20){\vector(1,0){18.5}}
\put(143.75,21){\makebox(0,0)[b]{\scriptsize$\equiv$}}

\put(155,0){\framebox(55,40){\scriptsize$\begin{aligned}&0.85\,\ketbra{\zap}{\zap}+\\&0.15\,\ketbra{\zbp}{\zbp}\end{aligned}$}}
\put(155,41){\makebox(0,0)[bl]{\scriptsize(c)}}

\put(182.5,-2){\vector(0,-1){18.5}}
\put(183.5,-11.25){\makebox(0,0)[l]{\scriptsize$-Q''<Q'$}}

\put(155,-62.5){\framebox(55,6){\scriptsize$\ketbra{\zbp}{\zbp}$}}
\put(155,-56.5){\framebox(55,34){\scriptsize$\ketbra{\zap}{\zap}$}}
\put(155,-21.5){\makebox(0,0)[bl]{\scriptsize(d)}}

%\put(182.5,-59.5){\line(0,-1){6}}\put(182.5,-59.5){\circle*{2}}
%\put(182.5,-65.5){\makebox(0,0)[t]{\scriptsize$\begin{aligned}&0.07\,\ketbra{\zbup}{\zbup}+\\&0.07\,\ketbra{\zbvp}{\zbvp}\end{aligned}$}}

\put(153,-42.5){\vector(-1,0){18.5}}

\put(77.5,-42.5){\framebox(55,20){\scriptsize$\ketbra{\zzp}{\zzp}$}}
\put(77.5,-62.5){\framebox(55,20){\scriptsize$\ketbra{\zzp}{\zzp}$}}
\put(77.5,-21.5){\makebox(0,0)[bl]{\scriptsize(e)}}

\put(75.5,-42.5){\vector(-1,0){18.5}}

\put(0,-42.5){\framebox(55,20){\scriptsize$\ketbra{\zzp}{\zzp}$}}
\put(0,-62.5){\framebox(55,20){\scriptsize$\ketbra{\zxp}{\zxp}$}}
\put(0,-21.5){\makebox(0,0)[bl]{\scriptsize(f)${}\equiv{}$(a)}}

\put(27.5,-20.5){\vector(0,1){18.5}}
\put(26.5,-11.25){\makebox(0,0)[r]{\scriptsize${}\equiv{}$}}
%\put(75.5,-20.5){\vector(-1,1){18.5}}

%\put(75.5,-42.5){\vector(-1,0){18.5}}

%\put(0,-42.5){\framebox(55,20){\scriptsize$\begin{aligned}&0.5\,\ketbra{\zzup}{\zzup}+\\&0.5\,\ketbra{\zzvp}{\zzvp}\end{aligned}$}}
%\put(0,-62.5){\framebox(55,20){\scriptsize$\begin{aligned}&0.5\,\ketbra{\zxup}{\zxup}+\\&0.5\,\ketbra{\zxvp}{\zxvp}\end{aligned}$}}
%\put(0,-21.5){\makebox(0,0)[bl]{\scriptsize(f)}}

%\multiput(27.5,-20.5)(0,6){2}{\line(0,1){4}}
%\put(27.5,-8.5){\vector(0,1){6.5}}
%\put(28.5,-11.25){\makebox(0,0)[l]{\scriptsize($-Q'''>Q'+Q''$)}}
%\put(75,-26){\vector(-1,1){20}}
\end{picture}
\caption{Quantum gas experiment from Tatiana's point 
of view}\label{fig:alfred}
\end{figure}
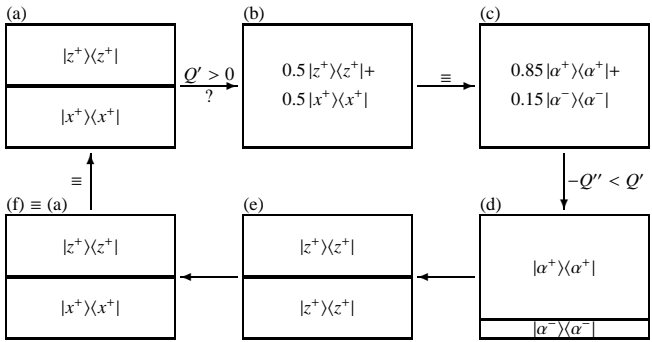

Peres' demonstration~\citep[pp.~275--277]{peres1995} can
be presented as follows. The physicist Tatiana describes
the internal quantum degrees of freedom of her ideal gases
by means of a spin\nbd-1/2 statistical-\bd matrix space
(with the related set of \POVM s). She starts
(Fig.~\ref{fig:alfred}, step a) with two quantum ideal
gases equally divided into two compartments having volumes
$V/2$ each and separated by an impermeable \diaphragm. In
the upper compartment there is a $\zz$-gas, in the lower a
$\zx$-gas, where $\zz$ and $\zx$ are the statistical
matrices defined in the previous section,
Eqs.~\eqref{eq:zplus} and~\eqref{eq:xplus}. Since they are
non-\bd orthogonal, $\tr(\zz\zx) \neq 0$, there are no
means to distinguish with certainty the two gases.
%% a particle of the $\zz$-gas from one of the $\zx$-gas
%% with certainty. 
%a preparation represented by $\zz$ from one represented by
%$\zx$.
%% \footnote{Note that this statement corresponds to
%% von~Neumann's and Peres'
%% `orthogonality${}\lmatimplies{}$distinguishability'
%% \eqref{eq:first}, and \emph{not} to our
%% `distinguishability${}\lmatimplies{}$orthogonality'
%% \eqref{eq:firstbis}.}
This implies for Tatiana the non-existence of two
semi-permeable \diaphragms\ with the property of being,
the one, completely transparent to the $\zz$-gas and
completely opaque to the $\zx$-gas, and vice versa for the
other, as discussed in the previous section.

Enters a ``wily inventor'', as Peres calls
him~\citep[p.~275]{peres1995}; let us call him Willard. He
claims having produced two such semi-permeable \diaphragm
s, which can completely distinguish and separate the two
gases. By means of them he reversibly mixes the two gases,
\emph{obtaining} work equal to $Q'=\zN \zk \zT \ln 2
\approx 0.693\, \zN \zk \zT$, \cf\ 
Eq.~\eqref{eq:worksepar}.  We have now a single chamber of
volume $V$ filled with a half-half mixture of $\zz$- and
$\zx$-gases (step b).

From Tatiana's point of view, the situation is now the
same as that discussed in the last example of the previous
section: the gas mixture is equivalent (step d) to
%% a single $\zmix$-gas with
%% \begin{equation}
%% \label{eq:mixrho}
%% \zmix \defin \frac{1}{2} \ketbra{\zzp}{\zzp} + 
%% \frac{1}{2} \ketbra{\zxp}{\zxp} \corr
%% \frac{1}{4}\begin{pmatrix}
%% 3&1\\1&1
%% \end{pmatrix},
%% \end{equation}
%as discussed in the previous section, and she treats it as
a mixture of approximately $0.854$ parts of an $\za$-gas
and $0.146$ parts of an $\zb$-gas, where $\za$ and $\zb$
are the statistical matrices defined in
Eq.~\eqref{eq:newketalf}. Tatiana uses two semi-permeable
\diaphragms\ to separate the two $\zab$-gases, the
$\za$-gas into a $0.854$ fraction of the volume $V$, and
the $\zb$-gas into the remaining $0.146$ fraction (so that
they have the same pressure), and \emph{spends} work equal
to $-Q'' = -\zN \zk \zT (0.854 \ln 0.854 + 0.146 \ln
0.146)\approx 0.416\, \zN \zk \zT$, \cf\ Eq.~\eqref{eq:worknoondingsepar}.

Tatiana then performs two operations corresponding to
unitary rotations of the statistical matrices associated
to the two gases to a common one, say $\zz$, so that the
two compartments now contain for her the same $\zz$-gas;
she then eliminates the \diaphragms and reinserts another
impermeable one to divide the gas into two compartments of
equal volume (step e), and finally performs again an
operation represented by a rotation $\zz \mapsto \zx$ of
the statistical matrix associated to the gas in the lower
chamber. In this way she has apparently re-established the
original condition of the gases (step a), which have thus
undergone a cycle. These last operations are assumed to be
performable without expenditure or gain of work, hence
without heat exchange
either.\footnote{\label{fn:isochoric}Note that
von~Neumann~\citep[pp.~194 and 197]{vonneumann1932c} and
Peres~\citep[p.~275]{peres1995} assert that unitary
rotations can be realised by processes involving no
\emph{heat} exchange, but work exchange is allowed and
indeed sometimes necessary. However, in our present
discussion we have assumed all processes to be isothermal
and all gases ideal, and this implies that any
\emph{isochoric} exchange of work must be accompanied by
an equivalent exchange of heat (see
\sect~\ref{sec:matintro}); for this reason must Tatiana's
final isochoric unitary rotations be performed with no
energy exchange. This issue is related to the problematic
way in which the quantum and classical or thermodynamical
descriptions are combined; namely, the statistical
matrices are not thermodynamic variables \langfrench{au
pair} with the real numbers ($V$ and $T$) describing the
gas.}

Tatiana summarises the results as follows: the total
entropy change is naught because the initial and final
conditions are the same: $\incr S = 0$. The total heat
\emph{absorbed by the gases} equals the work \emph{done by
them} and amounts to
\begin{equation}
\label{eq:heatabscycle}
Q =Q' + Q'' \approx (0.693 - 0.416)\, \zN \zk \zT = 
0.277\, \zN \zk \zT > 0.
\end{equation}
Hence, we have a violation of the second
law~\eqref{eq:seclawcycl} because for Tatiana's gases
\begin{equation}
\label{eq:viollaw}
Q/\zT > 0 = \incr S.
\end{equation}

Tatiana accuses Willard of having violated the second law
by means of his strange semi-permeable \diaphragms\ that
``separate non-\bd orthogonal 
%statistical matrices
states''.

\section{Jaynes' demonstration\label{sec:jaynesdem}}

We leave the quantum laboratory where Tatiana and Willard
are now arguing after their experiment, and enter an
adjacent classical laboratory, where we shall look at
Jaynes' demonstration~\citep{jaynes1992}. The situation
here is in many respects very similar to the previous,
though it is \emph{completely ``classical''}.

\begin{figure}[t]
\setlength{\unitlength}{0.0047\columnwidth}
\begin{picture}(210,57)(0,-8)
\put(0,20){\framebox(55,20){\scriptsize Ar}}
\put(0,0){\framebox(55,20){\scriptsize Ar}}
\put(0,41){\makebox(0,0)[bl]{\scriptsize(a)}}

\put(57,20){\vector(1,0){18.5}}
\put(66.25,21){\makebox(0,0)[b]{\scriptsize$Q'>0$}}
\put(66.25,19){\makebox(0,0)[t]{\scriptsize?}}

\put(77.5,0){\framebox(55,40){\scriptsize Ar}}
\put(77.5,41){\makebox(0,0)[bl]{\scriptsize(b)}}

\put(134.5,20){\vector(1,0){18.5}}
%\put(143.75,21){\makebox(0,0)[b]{\scriptsize$\equiv$}}

\put(155,20){\framebox(55,20){\scriptsize Ar}}
\put(155,0){\framebox(55,20){\scriptsize Ar}}
\put(155,41){\makebox(0,0)[bl]{\scriptsize(c)${}\equiv{}$(a)}}

\put(182.5,-2){\line(0,-1){6}}
\put(182.5,-8){\line(-1,0){155}}
\put(27.5,-8){\vector(0,1){6}}
\put(105,-7){\makebox(0,0)[b]{\scriptsize${}\equiv{}$}}
\end{picture}
\caption{Classical gas experiment from Johann's point 
of view}\label{fig:johann}
\end{figure}
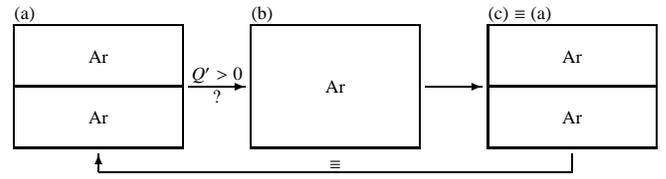

We have an ideal gas equally divided into two chambers of
volumes $V/2$ each and separated by an impermeable
\diaphragm\ (Fig.~\ref{fig:johann}, step a). For the
scientist Johann the gas in the two chambers is exactly
the same, say ``ideal argon'':\footnote{Real argon, of
course, behaves like an ideal gas only in certain ranges
of temperature and volume.} for him it would thus be
impossible, not to say meaningless, to find a
semi-permeable \diaphragm\ that be transparent to the gas
in the upper chamber and opaque to the gas in the lower
one, and another \diaphragm\ with the opposite properties.

The scientist Marie states nevertheless that she has in
fact two \diaphragms\ with those very properties. She uses
them to reversibly and isothermally mix the two halves of
the gas, \emph{obtaining} work equal to $Q' = \zN \zk \zT
\ln 2 \approx 0.693\, \zN \zk \zT$ (step b).

Yet, from Johann's point of view Marie has left things
exactly how they were: he just needs to reinsert the
impermeable \diaphragm\ in the middle of the vessel and
for him the situation is exactly the same as in the
beginning: \emph{the} gas is equally divided into two
chambers (steps c, a).

Johann's conclusion is the following: The initial and
final conditions of the gas are the same and so the total
entropy change vanishes: $\incr S = 0$. The work obtained
equals the heat absorbed by the gas,
\begin{equation}
\label{eq:heatabscyclejaynes}
Q = Q' \approx 0.693\, \zN \zk \zT > 0.
\end{equation}
The second law of thermodynamics states that this amount
of heat divided by the temperature cannot exceed the
entropy change,\footnote{Note that, strictly speaking,
ideal gases (or mixtures thereof) cannot undergo
irreversible processes (the usual example of ``free
expansion'' evidently assumes that the gas is not ideal
during the expansion).} $Q/\zT\le \incr S$, but this is
quite incompatible with Johann's conclusion that
\begin{equation}
Q/\zT > 0 = \incr S
%\tag*{(\ref{eq:seclawviolclas})$_\text{r}$}
\end{equation}
(\cf\ Tatiana's Eq.~\eqref{eq:viollaw}).

\medskip

Johann, however, is never dogmatic about his own knowledge
of the experimental facts. Asking Marie whether she is
able to reproduce her ``trick'' at will or whether it was
only chance, and upon her answer that the separation is
reproducible, he understands that where for him there was
only one gas there must actually be \emph{two different}
gases. This is indeed the case: Marie explains that the
two chambers initially contained two different kinds of
ideal-argon, of which Johann had no knowledge: argon~`$a$'
($\zAa$) and argon~`$b$' ($\zAb$). Argon~$a$ is soluble in
whafnium while argon~$b$ is not, but the latter is soluble
in whifnium, a property not shared by the
$a$~variety.\footnote{Jaynes explains that `whifnium', as
well as `whafnium', ``is one of the rare superkalic
elements; in fact, it is so rare that it has not yet been
discovered''~\citep[\sect5]{jaynes1992}.} Marie's
separation of the two gases $\zAa$ and $\zAb$ was possible
by means of two semi-permeable \diaphragms\ made of
whifnium and whafnium that take advantage of these
different properties.

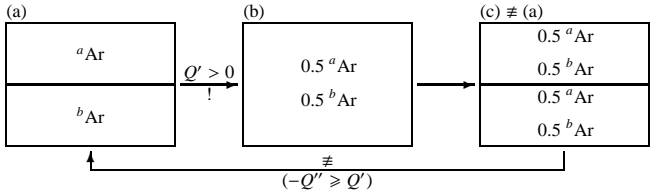
\begin{figure}[t]
\setlength{\unitlength}{0.0047\columnwidth}
\begin{picture}(210,57)(0,-8)%56,-11
\put(0,20){\framebox(55,20){\scriptsize ${}^a\text{Ar}$}}
\put(0,0){\framebox(55,20){\scriptsize${}^b\text{Ar}$}}
\put(0,41){\makebox(0,0)[bl]{\scriptsize(a)}}

\put(57,20){\vector(1,0){18.5}}
\put(66.25,21){\makebox(0,0)[b]{\scriptsize$Q'>0$}}
\put(66.25,19){\makebox(0,0)[t]{\scriptsize!}}

\put(77.5,0){\framebox(55,40){\scriptsize$\begin{aligned}&0.5\;{}^a\text{Ar}\\&0.5\;{}^b\text{Ar}\end{aligned}$}}
\put(77.5,41){\makebox(0,0)[bl]{\scriptsize(b)}}

\put(134.5,20){\vector(1,0){18.5}}
%\put(143.75,21){\makebox(0,0)[b]{\scriptsize$\equiv$}}

\put(155,20){\framebox(55,20){\scriptsize$\begin{aligned}&0.5\;{}^a\text{Ar}\\&0.5\;{}^b\text{Ar}\end{aligned}$}}
\put(155,0){\framebox(55,20){\scriptsize$\begin{aligned}&0.5\;{}^a\text{Ar}\\&0.5\;{}^b\text{Ar}\end{aligned}$}}
\put(155,41){\makebox(0,0)[bl]{\scriptsize(c)${}\nequiv{}$(a)}}

\put(182.5,-2){\line(0,-1){6}}
\put(182.5,-8){\line(-1,0){155}}
\put(27.5,-8){\vector(0,1){6}}
\put(105,-7){\makebox(0,0)[b]{\scriptsize${}\nequiv{}$}}
\put(105,-9){\makebox(0,0)[t]{\scriptsize($-Q'' \ge Q'$)}}

\end{picture}
\caption{Classical gas experiment from Marie's point 
of view}\label{fig:marie}
\end{figure}

We see (Marie's point of view, Fig.~\ref{fig:marie}) that
the second law is \emph{not} violated. Initially the two
gases $\zAa$ and $\zAb$ were completely separated in the
vessel's two chambers (step a). After mixing and
extracting work, the vessel contained an equal mixture of
$\zAa$ and $\zAb$ (step b). Upon Johann's reinsertion of
the impermeable \diaphragm\ the vessel is again divided in
two equal chambers, but each chamber contains a mixture of
$\zAa$ and $\zAb$ (step c), and this is \emph{different}
from the initial condition (step a): \emph{the cycle has
not been completed although it appeared so to Johann}, and
so the equation $\incr S = 0$ is not necessarily valid. To
close the cycle one has to use the semi-permeable
\diaphragms\ again to relegate the two gases to two
separate chambers, and must thereby \emph{spend} an amount
of work $-Q''$ at least equal to that previously obtained,
$-Q''\ge Q'$, and the second law~\eqref{eq:seclaw} for the
completed cycle is satisfied: $Q = Q' + Q'' \le 0 = \incr
S$.

The simple conclusion, drawn by
Jaynes~\citep[\sect3]{jaynes1992} in terms of entropy, is
that
\begin{quote}
it is necessary to decide at the outset of a problem which
macroscopic variables or degrees of freedom we shall
measure and/or control; and within the context of the
thermodynamic system thus defined, entropy will be some
function $S(X_1, \dotsc, X_n)$ of whatever variables we
have chosen. We can expect this to obey the second law
[$Q/\zT \le \incr S$] only as long as all experimental
manipulations are confined to that chosen set. If someone,
unknown to us, were to vary a macrovariable $X_{n+1}$
outside that set, he could produce what would appear to us
as a violation of the second law, since our entropy
function $S(X_1, \dotsc, X_n)$ might decrease
spontaneously, while his $S(X_1,\dotsc,X_n,X_{n+1})$
increases.
\end{quote}

This is old wisdom; for example, Grad had explained
thirty-one years earlier that~\citep[p.~325]{grad1961} (see
also~\citep{grad1952,grad1967})
\begin{quote}
the adoption of a new entropy is forced by the discovery
of new information. [\ldots] The existence of diffusion
between oxygen and nitrogen somewhere in a wind tunnel
will usually be of no interest. Therefore the
aerodynamicist uses an entropy which does not recognise
the separate existence of the two elements but only that
of ``air''. In other circumstances, the possibility of
diffusion between elements with a much smaller mass ratio
(\eg, 238/235) may be considered quite relevant.
\end{quote}

We can rephrase Grad's and Jaynes' remark shifting the
emphasis to the distinction between the experimental
situation and the mathematics which describes it: The fact
that some physicist can perform experimental operations
which contradict our mathematical description and which
apparently lead to violations of \eg\ the second law,
simply means that that physicist is able to control
physical phenomena which are not contemplated by our
mathematical description, and the second law is not
necessarily violated in that physicist's \emph{more
appropriate} mathematical description.

\section{Re-analysis of Peres' 
demonstration\label{sec:reanperes}}

With the insight provided by Grad and Jaynes, we can
return to the quantum laboratory and look with different
eyes at what happened there. If Willard can reproducibly
distinguish and separate with certainty the two physical
preparations represented by Tatiana through the non-\bd
orthogonal statistical matrices $\zz$ and $\zx$, this can
only have one meaning: these preparations \emph{have to}
be represented by \emph{orthogonal} statistical matrices
instead, at least in experimental situations in which
Willard takes advantage of his instrumental capabilities,
his ``tricks''. This is not in contradiction with
Tatiana's formalism: with the instruments and apparatus at
her disposal the two physical situations are not
distinguishable with certainty, and so the appropriate way
\emph{for her} of representing them was by non-\bd
orthogonal statistical matrices. But she can now share
Willard's instrumentation and knowledge and use an
accordingly more adequate mathematical description of the
physical facts.

We can imagine a possible explanation from Willard's point
of view (it is just a \emph{possible} one, and even more
drastic ones, requiring abandonment of the
quantum-mechanical formalism, might be necessary in other
instances). Willard explains that the internal quantum
degrees of freedom of the gases are best represented by a
spin-3/2 statistical-matrix space, of which Tatiana used a
subspace (more exactly, a projection) because of her
limited observational means; \ie, part of the
statistical-matrix space was ``traced out'' because Tatiana
used only instrumentation represented by a portion of the
total \POVM\ space. For instance, denoting by $\set{%
\ket{\zzup}, \ket{\zzum}, \ket{\zzvp}, \ket{\zzvm}
%\zuk, \zdk, \ztk, \zqk
}$ the basis for the Hilbert space used by Willard, with
$\braket{\zzup}{\zzvp} = \braket{\zzum}{\zzvm} \equiv 0$,
Tatiana could not distinguish, amongst others, the
preparations corresponding to
% $\zuk$ and to $\bigl(\ztk+\zqk\bigr)/\sqrt{2}$
$\ket{\zzup}$ and to $\ket{\zzvp}$, both of which she
represented as $\zzk$, nor those corresponding to
% $\ztk$ and to $\bigl(\zuk+\zdk\bigr)/\sqrt{2}$
$\ket{\zxup} \defin \bigl(\ket{\zzup} +
\ket{\zzum}\bigr)/\sqrt{2}$ and to $\ket{\zxvp} \defin
\bigl(\ket{\zzvp} + \ket{\zzvm}\bigr)/\sqrt{2}$, which she
denoted as $\zxk$. The projection is thus
\begin{gather}
\label{eq:proj}
\begin{aligned}
%\zuk 
\ket{\zzup}&\mapsto \zzk,&
%\zdk 
\ket{\zzum}&\mapsto \ket{\zzm},&
%\ztk 
\ket{\zzvp}&\mapsto \zzk,&
%\zuk 
\ket{\zzvm}&\mapsto \ket{\zzm},
\end{aligned}\\
\intertext{from which also follows}
\begin{aligned}
%\zuk 
\ket{\zxup}&\mapsto \zxk,&
%\zdk 
\ket{\zxum}&\mapsto \ket{\zxm},&
%\ztk 
\ket{\zxvp}&\mapsto \zxk,&
%\zuk 
\ket{\zxvm}&\mapsto \ket{\zxm},
\end{aligned}
\end{gather}
%% It is convenient to rename Willard's basis as
%% $\ket{\zzup} \equiv \zuk$, 
%% $\ket{\zzum} \equiv \zdk$, $\ket{\zxvp}
%% \equiv \ztk$, $\ket{\zxvm} \equiv\ \zqk$, 
which makes it evident that Tatiana cannot distinguish
preparations and experiments represented by the vectors
with a tilde from the corresponding ones represented by
accented vectors.\footnote{An even more evident notation
would be $\ket{\zzup} \equiv \ket{\zzp,\zzp}$,
$\ket{\zzvp} \equiv \ket{\zzp,\zzm}$, \etc, but
it might be misleading in other respects.}
%make one think of two ``separate systems

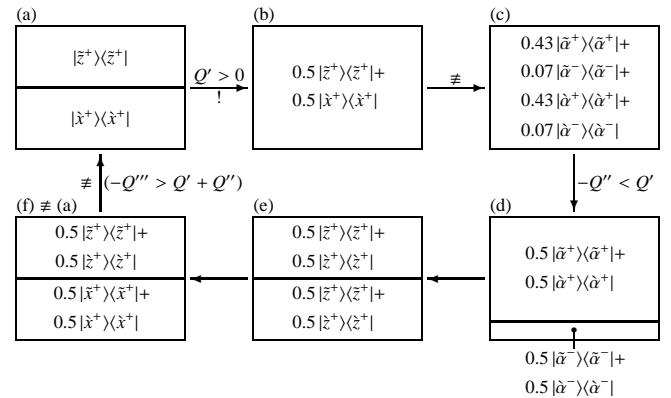
\begin{figure}[b]
\setlength{\unitlength}{0.0047\columnwidth}
\begin{picture}(210,129)(0,-82.5)
\put(0,20){\framebox(55,20){\scriptsize$\ketbra{\zzup}{\zzup}$}}
\put(0,0){\framebox(55,20){\scriptsize$\ketbra{\zxvp}{\zxvp}$}}
\put(0,41){\makebox(0,0)[bl]{\scriptsize(a)}}

\put(57,20){\vector(1,0){18.5}}
\put(66.25,21){\makebox(0,0)[b]{\scriptsize$Q'>0$}}
\put(66.25,19){\makebox(0,0)[t]{\scriptsize!}}

\put(77.5,0){\framebox(55,40){\scriptsize$\begin{aligned}&0.5\,\ketbra{\zzup}{\zzup}+\\&0.5\,\ketbra{\zxvp}{\zxvp}\end{aligned}$}}
\put(77.5,41){\makebox(0,0)[bl]{\scriptsize(b)}}

\put(134.5,20){\vector(1,0){18.5}}
\put(143.75,21){\makebox(0,0)[b]{\scriptsize$\nequiv$}}

\put(155,0){\framebox(55,40){\scriptsize$\begin{aligned}&0.43\,\ketbra{\zaup}{\zaup}+\\&0.07\,\ketbra{\zbup}{\zbup}+\\&0.43\,\ketbra{\zavp}{\zavp}+\\&0.07\,\ketbra{\zbvp}{\zbvp}\end{aligned}$}}
\put(155,41){\makebox(0,0)[bl]{\scriptsize(c)}}

\put(182.5,-2){\vector(0,-1){18.5}}
\put(183.5,-11.25){\makebox(0,0)[l]{\scriptsize$-Q''<Q'$}}

\put(155,-62.5){\framebox(55,6){}}
\put(155,-56.5){\framebox(55,34){\scriptsize$\begin{aligned}&0.5\,\ketbra{\zaup}{\zaup}+\\&0.5\,\ketbra{\zavp}{\zavp}\end{aligned}$}}
\put(155,-21.5){\makebox(0,0)[bl]{\scriptsize(d)}}

\put(182.5,-59.5){\line(0,-1){6}}\put(182.5,-59.5){\circle*{2}}
\put(182.5,-65.5){\makebox(0,0)[t]{\scriptsize$\begin{aligned}&0.5\,\ketbra{\zbup}{\zbup}+\\&0.5\,\ketbra{\zbvp}{\zbvp}\end{aligned}$}}

\put(153,-42.5){\vector(-1,0){18.5}}

\put(77.5,-42.5){\framebox(55,20){\scriptsize$\begin{aligned}&0.5\,\ketbra{\zzup}{\zzup}+\\&0.5\,\ketbra{\zzvp}{\zzvp}\end{aligned}$}}
\put(77.5,-62.5){\framebox(55,20){\scriptsize$\begin{aligned}&0.5\,\ketbra{\zzup}{\zzup}+\\&0.5\,\ketbra{\zzvp}{\zzvp}\end{aligned}$}}
\put(77.5,-21.5){\makebox(0,0)[bl]{\scriptsize(e)}}

\put(75.5,-42.5){\vector(-1,0){18.5}}

\put(0,-42.5){\framebox(55,20){\scriptsize$\begin{aligned}&0.5\,\ketbra{\zzup}{\zzup}+\\&0.5\,\ketbra{\zzvp}{\zzvp}\end{aligned}$}}
\put(0,-62.5){\framebox(55,20){\scriptsize$\begin{aligned}&0.5\,\ketbra{\zxup}{\zxup}+\\&0.5\,\ketbra{\zxvp}{\zxvp}\end{aligned}$}}
\put(0,-21.5){\makebox(0,0)[bl]{\scriptsize(f)${}\nequiv{}$(a)}}

\put(27.5,-20.5){\vector(0,1){18.5}}

%\multiput(27.5,-20.5)(0,6){2}{\line(0,1){4}}
%\put(27.5,-8.5){\vector(0,1){6.5}}
\put(26.5,-11.25){\makebox(0,0)[r]{\scriptsize${}\nequiv{}$}}
\put(28.5,-11.25){\makebox(0,0)[l]{\scriptsize($-Q'''>Q'+Q''$)}}

\end{picture}
\caption{Quantum gas experiment from Willard's point 
of view}\label{fig:willard}
\end{figure}

Thus, the preparations which Tatiana represents by $\zz$
and $\zx$ because for her they were indistinguishable, are
instead represented by Willard by
\begin{align}
\zzu &\defin \ketbra{\zzup}{\zzup} \corr
\left(\begin{smallmatrix}
1&0&0&0\\ 0&0&0&0\\ 0&0&0&0\\ 0&0&0&0
\end{smallmatrix}\right), &
\zxv &\defin \ketbra{\zxvp}{\zxvp} \corr
\left(\begin{smallmatrix}
0&0&0&0\\ 0&0&0&0\\ 0&0&1&0\\ 0&0&0&0
\end{smallmatrix}\right),
\end{align}
which are clearly orthogonal, $\tr(\zzu\zxv) = 0$, because
for him the two corresponding preparations are
distinguishable. From his point of view, the process went
as follows. The two compartments initially contained
$\zzu$- and $\zxv$-gases (Fig.~\ref{fig:willard}, step a).
With his semi-permeable \diaphragms\ he mixed the two
distinguishable gases with extraction of work (step b), so
that the chamber eventually contained a $\zmi$-gas with
\begin{equation}
\label{eq:mixwill}
%% \zmi = \frac{1}{2} \ketbra{\zzup}{\zzup} +
%% \frac{1}{2} \ketbra{\zxvp}{\zxvp}.
\zmi = \frac{1}{2} \zzu %\ketbra{\zzup}{\zzup} 
+ \frac{1}{2} \zxv %\ketbra{\zxvp}{\zxvp} 
\corr \tfrac{1}{2}\left(\begin{smallmatrix}
1&0&0&0\\ 0&0&0&0\\ 0&0&1&0\\ 0&0&0&0
\end{smallmatrix}\right).
\end{equation}

Tatiana's subsequent separation by means of her
semi-permeable \diaphragms\ (steps c, d), distinguishing
the preparations corresponding to 
$\za$ and $\zb$,
%% \footnote{We shall see in a moment that this step is
%% indeed \emph{irreversible}.}
is represented by Willard by the \POVM
\begin{gather}
\begin{split}
\zwEF &\defin \ketbra{\zwbBk}{\zwbBk} +
\ketbra{\zwcCk}{\zwcCk},\\
 &\corr \tfrac{1}{4}\left(\begin{smallmatrix}
2\pm\sqrt{2}&\pm\sqrt{2}&0&0\\
\pm\sqrt{2}& 2\mp \sqrt{2}&0&0 \\
0&0&2\pm\sqrt{2}&\pm\sqrt{2}\\
0&0&\pm\sqrt{2}& 2\mp \sqrt{2} \\
\end{smallmatrix}\right),
\end{split}
\\
\intertext{with}
%\begin{split}
\ket{\zwbB} \defin \Bigl(2\pm
\sqrt{2}\Bigr)^{-\frac{1}{2}} \bigl(\ket{\zzupm} \pm
\ket{\zxupm}\bigr),
%% \\
%% &\equiv \frac{1}{2}
%% \Bigl[\pm\Bigl(2\pm\sqrt{2}\Bigr)^{\frac{1}{2}}
%% \ket{\zup} + \Bigl(2\mp\sqrt{2}\Bigr)^{\frac{1}{2}}
%% \ket{\zdp}\Bigr],
%% \end{split}
\\
%\begin{split}
\ket{\zwcC} \defin \Bigl(2\pm
\sqrt{2}\Bigr)^{-\frac{1}{2}} \bigl(\ket{\zzvpm} \pm
\ket{\zxvpm}\bigr),
%% \\
%% &\equiv \frac{1}{2}
%% \Bigl[\pm\Bigl(2\pm\sqrt{2}\Bigr)^{\frac{1}{2}}
%% \ket{\ztp} + \Bigl(2\mp\sqrt{2}\Bigr)^{\frac{1}{2}}
%% \ket{\zqp}\Bigr]
%% \end{split}
\end{gather}
\cf\ Eq.~\eqref{eq:newketalf}, and the associated \CPM s
$\zmi \mapsto \zwEF\zmi\zwEF/ \tr( \zwEF\zmi\zwEF)$.

That separation led to a compartment, with volume $0.854
V$, containing the gas \emph{mixture}
\begin{equation}
\label{eq:mixaz}
\frac{1}{2} \ketbra{\zaup}{\zaup}+
\frac{1}{2} \ketbra{\zavp}{\zavp},
\end{equation}
and the other compartment, with volume $0.146 V$,
containing the gas \emph{mixture}
\begin{equation}
\label{eq:mixbz}
\frac{1}{2} \ketbra{\zbup}{\zbup}+
\frac{1}{2} \ketbra{\zbvp}{\zbvp}.
\end{equation}
Tatiana could not perceive that these were mixtures,
because of her limited instrumentation.

The following step corresponded to the rotations
\begin{gather}
\ketbra{\zwbBpk}{\zwbBpk} \mapsto \ketbra{\zzup}{\zzup},
\qquad\ketbra{\zwcCpk}{\zwcCpk} \mapsto
\ketbra{\zzvp}{\zzvp}\\
% \defin\bigl(\ket{\zxvp}+\ket{\zqp}\bigr)/\sqrt{2}
\intertext{for gases in the upper compartment, and}
\ketbra{\zwbBmk}{\zwbBmk} \mapsto \ketbra{\zzup}{\zzup},
\qquad \ketbra{\zwcCmk}{\zwcCmk} \mapsto
\ketbra{\zzvp}{\zzvp}
% \defin \bigl(\ket{\zxvp}+\ket{\zqp}\bigr)/\sqrt{2}
\end{gather}
for the gases in the lower compartment (remember that
\mbox{$\braket{\zzup}{\zzvp}=0$}). The successive
elimination and reinsertion of the impermeable \diaphragm\ 
led to two compartments of equal volumes $V/2$ and equal
content, \viz\ the equal \emph{mixture} of
$\ketbra{\zzup}{\zzup}$- and $\ketbra{\zzvp}{\zzvp}$-gases
(step e).

The final rotation for the gas in the lower compartment,
\begin{equation}
\ket{\zzup} \mapsto \ket{\zxup},\qquad 
%\defin \bigl(\ket{\zzup}+\ket{\zdp}\bigr)/\sqrt{2} 
\ket{\zzvp} \mapsto \ket{\zxvp}
\end{equation}
(remember that \mbox{$\braket{\zxup}{\zxvp}=0$}), only led
to two equal compartments containing the \emph{mixtures}
of $\tfrac{1}{2} \ketbra{\zzup}{\zzup}+ \tfrac{1}{2}
\ketbra{\zzvp}{\zzvp}$ and $\tfrac{1}{2}
\ketbra{\zxup}{\zxup}+ \tfrac{1}{2} \ketbra{\zxvp}{\zxvp}$
gases respectively (step f). This is of course
\emph{different} from the initial situation (step a); but
for Tatiana, whose instrumentation was limited with
respect to Willard's, the initial and final conditions
appeared identical.

The second law is \emph{not} violated, because \emph{the
cycle has not been completed}, and the equation $\incr S =
0$ does thus not necessarily hold. It is easy to see that
in order to return to the initial condition an amount of
work $-Q''' \ge 4\times (1/4) \zN \zk \zT \ln 2 \approx
0.693\, \zN \zk \zT$ has to be \emph{spent} to separate
the $\ketbra{\zzup}{\zzup}$-gas from the
$\ketbra{\zzvp}{\zzvp}$-gas, and analogously for the
$\ketbra{\zxup}{\zxup}$- and
$\ketbra{\zxvp}{\zxvp}$-gases. A final operation must then
be performed corresponding to the rotations of the
statistical matrices $\ketbra{\zzvp}{\zzvp}$ and
$\ketbra{\zxup}{\zxup}$ to $\ketbra{\zzup}{\zzup}$ and
$\ketbra{\zxvp}{\zxvp}$ respectively, and we have finally
reached again the initial condition (step a). The total
amount of heat \emph{absorbed} by the gases, corresponding
to the work performed on them would then be
\begin{multline}\label{eq:totalabsperes}
Q =Q' + Q'' + Q''' \le
%\approx{} 
(0.693 - 0.416 - 0.693) \zN
\zk \zT = {}\\
-0.416 \zN \zk \zT \le 0 = \incr S,
\end{multline}
and the second law, for the completed cycle, is satisfied
(strictly so: we see that the whole process is
irreversible, and it is easy to check that the only
irreversible step was the separation performed by Tatiana
into $\za$- and $\zb$-gases).

\section{Conclusion}
\label{sec:concldisc}

The re-analysis, with Jaynes' (and Grad's) insight, of the
simple quantum experiment which seemed to violate the
second law leads to the following almost trivial
conclusion: if the physicist Willard can reproducibly
distinguish two physical preparations that the physicist
Tatiana represents by non-\bd orthogonal statistical
matrices, then no necessary violation of the second law of
thermodynamics is implied from Willard's point of view. On
the other hand it is certain that the particular
statistical-matrix space adopted by Tatiana for the
phenomenon's description is not (any longer) adequate, and
has to be amended (in extreme cases a non-\bd quantum-\bd
mechanical description might be necessary) in order to
avoid inconsistencies like \eg\ seeming violations of the
second law. Alternatively, Tatiana can keep on using the
unamended mathematical description, but she must then
renounce to treat with it situations involving the new
experimental possibility and the related phenomena, in
order to avoid inconsistencies.
% like \eg\ false violations of the second law.
% Some further considerations may be made regarding this
% conclusion.

This conclusion emphasises the distinction, somewhat
obscured in von~Neumann's statements but often stressed
by Peres~\citep{peres1984,peres1995,peres2000a,peres2002}
and Jaynes~\citep{jaynes2003} amongst
others,\footnote{\Eg\ 
Ekstein~\citep{ekstein1967,ekstein1969},
Giles~\citep{giles1968,giles1970,giles1979}, Foulis and
Randall~\citep{foulisetal1972a,randalletal1973b,foulisetal1978},
Band and Park~\citep{parketal1976,bandetal1976}; \cf\ also
References~\citep{ballentine1970,hardy2001,mana2004b}.}
between physical phenomena and their mathematical
description: ``quantum phenomena \emph{do not} occur in a
Hilbert space. They occur in a
laboratory''~\citep{peres2002}.\footnote{A similar
explicit observation had already been made by
%% I think we can
%% find early explicit expressions of this distinction in
%% papers by Einstein \etal~\citep{einsteinetal1935},
%% Fano~\citep{fano1957}, 
%% Stapp~\citep{stapp1971,stapp1972},
%% Giles~\citep{giles1970}, Foulis and
%% Randall~\citep{randalletal1970,foulisetal1972a},
%% Ballentine~\citep{ballentine1970}, and
%% especially 
Band and Park~\citep{parketal1976%,bandetal1976
}: ``experimenters do not apprehend Hilbert vectors; they
gather \emph{numerical data}'' (their emphasis).} (The
present author is in fact guilty of unclarity about this
important distinction in a previous
paper~\citep{mana2003}.) The usual metonymic
expression\footnote{Metonymy is ``a figure of speech
consisting of the use of the name of one thing for that of
another of which it is an attribute or with which it is
associated (as `crown' in `lands belonging to the
crown')'' (Merriam-Webster Online Dictionary); it
perfectly applies to our case.} ``to distinguish two
statistical matrices'' is certainly handy, but must be
used with a grain of salt: what we distinguish is in fact
two physical situations, facts, phenomena, or
preparations; not two statistical matrices. The latter
should mathematically reflect what we can do with these
preparations, \eg whether we can distinguish them, and be
amended whenever new experimental facts
appear.\footnote{We must not forget, however, that the
mathematical formalism is often expanded with no or very
little experimental input and successfully leads thus to
new discoveries (think of \eg\ general relativity, Dirac's
positron, or the Clausius-Duhem inequality).}
%% the statistical matrix is only
%% a mathematical object used to describe the statistical
%% properties of these phenomena.
%% the exact definition, or better,
%% delimitation
%% %circumscription
%% of a given collection of physical phenomena and the
%% correct assignment to these of appropriate statistical
%% matrices, \POVM s, and \CPM s, is always a delicate
%% matter.
%% The curtailed way of speaking that so often
%% appears in the literature suggests a sort of intimate and
%% exclusive relation between a `preparation' and the statistical
%% matrix which describes it;

\begin{acknowledgements}
I thank Gunnar Bj\"ork and Anders M\aa{}nsson for many
discussions regarding von~Neumann's demonstration, Gunnar
also for encouragement and for precious comments on the
paper, Louise and Anna for encouragement, and Louise for
invaluable odradek.
\end{acknowledgements}

%\appendix

\providecommand{\href}[2]{#2}
\providecommand{\eprint}[2][http://arxiv.org/abs/]{\href{#1#2}{\texttt{#2}}}
\renewcommand{\eprint}[2][http://arxiv.org/abs/]{\href{#1#2}{arxiv.org/#2}}
\newcommand{\arxiveprint}[1]{%\url{arXiv.org} 
%eprint
\eprint{#1}}
\newcommand{\citein}[1]{\textnormal{\citet{#1}}}

%\bibliography{bibliography,blocal}\end{document}

\end{document}